\documentclass[10pt]{article}
\usepackage{epsf}
\usepackage{amsmath}

\usepackage{amssymb}
\usepackage{cite}
\usepackage{url}

\usepackage{graphicx}
\usepackage{xcolor}

\newenvironment{refone}{\begingroup\color{black}}{\endgroup}
\newenvironment{reftwo}{\begingroup\color{black}}{\endgroup}
\newenvironment{refthree}{\begingroup\color{black}}{\endgroup}
\newenvironment{pubrev}{\begingroup\color{black}}{\endgroup}

\begin{document}

\begin{center}
\noindent
{\bf \Large Consistency Conditions for Astrons\\}

\vspace{0.5cm}  

\vspace{1.cm}

\noindent
{  $^{1,2}$Claudio Corian\`o and $^1$Leonardo Torcellini}

\vspace{0.5in}

\noindent
{\it   $^1$Dipartimento di Matematica e Fisica "Ennio De Giorgi",\\ 
Universit\`{a} del Salento and INFN-Lecce,\\ Via Arnesano, 73100 Lecce, Italy.}\\
{\it  $^2$CNR nanotec, Lecce \\}

\vspace{0.5in}

\begin{abstract}
\noindent
{ We reassess the consistency conditions for a hypothetical population
of primordial, electrically charged compact objects, here called astrons.
This paper formulates astrons not as a completed model, but as a set of
quantitative consistency tests for a primordial large-charge compact-object
scenario.    The
fiducial phenomenological parameters,
\(M_A\sim10^{12}M_\odot\), \(Q_A\sim4\times10^{32}\,\mathrm{C}\), and
megaparsec-scale separations, are treated as a benchmark to be tested rather
than as an established outcome of the model.  We show explicitly that ordinary
accretion-driven charge separation produces charges only of order
\(10^{11}\)--\(10^{14}\,\mathrm C\) for this mass, leaving an
18--21 order-of-magnitude gap relative to the fiducial branch.  Thus the
large-charge branch, if it exists, must arise from a primordial
charge-concentration mechanism not supplied by the minimal capture model.
Likewise, the failure of linear Debye--H\"uckel screening in the enormous
electrostatic potential of an astron does not demonstrate charge survival; it
identifies a nonlinear neutralization and kinetic-transport problem.  Finally,
a cosmologically relevant abundance of \(10^{12}M_\odot\) compact objects
requires mean separations of several megaparsecs and must satisfy discreteness,
Poisson-power, clustering, and dynamical constraints.  The homogeneous Coulomb
interaction energy scales as \(a^{-4}\), so it cannot by itself act as a
late-time cosmological constant.  \begin{refone}Moreover, the Lorentz force
acts directly only on charged astrons, not on the no-net-charge population of
galaxies, photons and neutral dark matter.  Any viable cosmological
implementation must therefore derive a metric-level or domain-averaged
acceleration shared by the neutral cosmic flow, rather than merely a pairwise
repulsion among charged sources.\end{refone} }
\end{abstract}
\begin{pubrev}
\begin{flushleft}
\end{flushleft}
\end{pubrev}
\end{center}

\newpage

\tableofcontents

\newpage

\section{Introduction}

\noindent
The origin of the accelerated expansion of the Universe remains one of the central
open problems in modern cosmology. Within the standard $\Lambda$CDM framework,
the observed late-time acceleration is encoded in a cosmological constant, yet the
microscopic origin of that term remains obscure. This motivates the study of alternative
scenarios in which long-range interactions or collective dynamics modify cosmic
expansion without introducing a fundamental vacuum component.

In this paper we revisit a framework in which the relevant degrees of freedom are
primordial, electrically charged compact objects carrying charges of common sign.
The present construction is closely related to earlier ideas on cosmological
scenarios driven by extremely massive charged compact objects, including the electromagnetic
accelerating-universe proposal and the later unified dark-matter/dark-energy model
\cite{Frampton2022EAU,Frampton2023DMDE}.

Following earlier work on Primordial Extremely
Massive Naked Singularities (PEMNS), we refer to these objects simply as \emph{astrons}.
The fiducial parameters usually associated with this scenario are masses
$M_A \sim 10^{12} M_\odot$, charges $|Q_A| \sim 10^{32}\,\mathrm{C}$, and typical
separations of order megaparsecs. If such objects exist and retain their charge
over cosmological times, their mutual Coulomb interaction could in principle influence
large-scale dynamics.

{\begin{refone}
There is, however, a distinction that must be kept explicit from the beginning.
The Lorentz force acts directly only on matter with nonzero charge-to-mass ratio.
It can accelerate charged astrons relative to one another, but it does not
directly push a no-net-charge population of galaxies, photons and neutral dark
matter.  The rest of the Universe can participate in an astron-driven expansion
only through the geometry sourced by the total Einstein--Maxwell stress tensor,
through gravitational binding to charged structures, or through a genuine
inhomogeneous backreaction effect.  This distinction is central to the
cosmological assessment below.
\end{refone}}

The conventional astrophysical expectation is that macroscopic black holes are
effectively described by their mass and spin, with any residual electric charge
often assumed to be negligible in realistic environments \cite{Bambi2019}. At
the same time, several recent analyses have emphasized that small but non-zero
charges can be generated by astrophysical processes and need not be irrelevant:
even charges far below extremality can influence the motion of charged particles,
shift orbital structures, and modify electromagnetic observables near compact
objects \cite{ZajacekTursunov2019,ZajacekEtAl2018}. More generally, long-range
interactions are known to have potentially large astrophysical consequences and
are tightly constrained in other sectors, including dark-matter interactions with
the Standard Model \cite{BogoradGrahamRamani2025}.

The present analysis is deliberately narrower than a full replacement of
$\Lambda$CDM. The relevant issue is whether the astron picture is internally
consistent enough to warrant further study. Three questions are central. First,
can a collapsing object acquire macroscopic charge quickly enough for charging
to matter during formation? Second, once produced, can such a charge persist
against accretion, plasma neutralization, and quantum discharge? Third, under
what plasma conditions does Debye screening invalidate or constrain the
long-range force picture?

These questions matter because each one can obstruct the entire scenario. Screening
in the intergalactic medium could suppress long-range Coulomb forces, while accretion
or pair production could eliminate any large charge soon after formation. Large charge-to-mass
ratios also raise geometric issues, since super-extremal Reissner--Nordstr\"om
configurations do not possess an event horizon. Before confronting observations,
one must therefore determine which elements of the framework survive a minimal consistency
analysis.
\subsection{Picturing an astron}
For this astrophysical body the scale \(GM_A/c^2\) should not be interpreted as a stellar photospheric radius.
It is a relativistic compactness scale. This distinction is important: for a
neutral Schwarzschild object, a material surface at or below \(2GM_A/c^2\) would
lie at the black-hole threshold. For a charged astron the corresponding statement
is more subtle, because the exterior horizon condition is the
Reissner--Nordstr\"om condition controlled by \(\Xi\), defined in
Eq.~\eqref{eq:xi_eta_def}. Thus the gravitational radius is used here mainly as
a reference length for compactness and for order-of-magnitude estimates, not as a
claim that the object has a luminous surface at that radius. In that sense, an
astron is not expected to resemble an ordinary star, whose visible emission is
produced by thermal radiation from a material outer layer. If the astron is instead
modeled as an ultra-compact charged object or naked singularity, it need not possess
any such radiating surface at all. Consequently, a large relativistic scale does
not by itself imply that the astron would appear as a bright optical source.
Significant electromagnetic emission would arise only if the object were surrounded
by accreting gas or by a sufficiently dense plasma capable of producing radiation.
In the absence of such an environment, the astron could remain electromagnetically
dim and would be more naturally probed through its gravitational influence, lensing
properties, or interactions with ambient matter rather than through direct
stellar-like light. For \(M_A\sim10^{12}M_\odot\), the reference angular scale
associated with the compactness length is of order
\(R_{\rm ref}\sim GM_A/c^2\sim 0.05\,\mathrm{pc}\), or
\(R_{\rm ref}\sim 0.1\,\mathrm{pc}\) if one uses \(2GM_A/c^2\). The corresponding
angular scale is
\begin{equation}
\theta_A \simeq 206265\,\frac{R_{\rm ref}(\mathrm{pc})}{D(\mathrm{pc})}\ \mathrm{arcsec}
\simeq 0.01''
\left(\frac{R_{\rm ref}}{0.05\,\mathrm{pc}}\right)
\left(\frac{D}{1\,\mathrm{Mpc}}\right)^{-1}.
\end{equation}
Thus even at a distance of \(1\,\mathrm{Mpc}\) this reference angular scale is
only of order \(10^{-2}\) arcsec, while at \(5\,\mathrm{Mpc}\) it falls to a few
milliarcseconds.
An astron would therefore not generically appear as an obvious resolved luminous
disk in the sky.

It is also useful to compare this fiducial compactness scale with more familiar
astrophysical systems. For \(M_A\sim10^{12}M_\odot\), the gravitational radius is
\begin{equation}
r_g=\frac{GM_A}{c^2}\simeq 1.5\times10^{15}\,\mathrm{m}
\simeq 10^4\,\mathrm{AU}\simeq 0.05\,\mathrm{pc},
\end{equation}
while the corresponding Schwarzschild radius is
\begin{equation}
r_s=2r_g\simeq 2\times10^4\,\mathrm{AU}\simeq 0.1\,\mathrm{pc}.
\end{equation}
This is vastly larger than the planetary Solar System: Pluto orbits at a mean
distance of about \(40\,\mathrm{AU}\), and the heliopause lies only at roughly
\(10^2\,\mathrm{AU}\). At the same time, an astron remains small on galactic
length scales. A sub-parsec compactness length is tiny compared with a galactic
disk, whose size is typically measured in kiloparsecs. Relative to known
supermassive black holes, however, the fiducial astron scale is exceptionally
large. The Schwarzschild
radius of Sgr A*, with \(M\simeq4\times10^6M_\odot\), is only about
\(0.08\,\mathrm{AU}\), while that of M87*, with
\(M\simeq6.5\times10^9M_\odot\), is about \(10^2\,\mathrm{AU}\). Thus an astron
with \(M_A\sim10^{12}M_\odot\) would have a relativistic length scale hundreds
to thousands of times larger than that of ordinary supermassive black holes,
while still occupying only a sub-parsec region inside a galaxy. This comparison
again refers to the compactness scale, not necessarily to a material surface. If
a neutral object had \(R\lesssim r_s\), it would be a black hole; in the charged
case the question is instead whether the exterior lies below, at, or above the RN
extremality bound. The physical radius or interior completion of a horizonless or
singular astron is therefore model dependent.

\subsection{Primordial timing and status of the benchmark}
\label{subsec:primordial_status}

The fiducial large charge should not be interpreted as the natural endpoint of
ordinary late-time accretion.  In the minimal accretion picture discussed below,
electrostatic feedback quickly limits the charge to a much smaller
saturation value.  Thus, if the large branch
\(M_A\sim10^{12}M_\odot\), \(Q_A\sim4\times10^{32}\,{\rm C}\) exists, it should
be regarded as a primordial or very early-universe charge-concentration
hypothesis.  The relevant question is not simply whether an already formed
galaxy can accrete enough charge, but whether a rare primordial object can be
born with a small charge imbalance that later survives plasma neutralization.

A useful way to locate such an object in cosmic history is to compare its mass
with the horizon mass in the radiation era.  For a radiation-dominated universe,
up to factors of order unity,
\begin{equation}
H\simeq \frac{1}{2t},\qquad
\rho\simeq \frac{3H^2}{8\pi G}\simeq \frac{3}{32\pi Gt^2}.
\end{equation}
Taking the causal length to be \(R_H\sim 2ct\), the mass inside the horizon is
\begin{equation}
M_H(t)\sim \frac{4\pi}{3}\rho R_H^3
\sim \frac{c^3t}{G}.
\label{eq:horizon_mass_estimate}
\end{equation}
This is the standard order-of-magnitude estimate used in discussions of
primordial black-hole formation \cite{CarrHawking1974}.  Numerically,
\begin{equation}
M_H(t)\simeq
2\times10^5M_\odot
\left(\frac{t}{1\,{\rm s}}\right),
\end{equation}
or equivalently
\begin{equation}
t_A\sim
\frac{G M_A}{c^3}
\simeq
5\times10^6\,{\rm s}
\left(\frac{M_A}{10^{12}M_\odot}\right).
\end{equation}
Thus the fiducial astron mass corresponds, at this crude horizon-mass level, to
a time of order weeks to months after the Big Bang.  This estimate is only a
causal mass scale.  It is not a formation theory, and it does not explain the
charge.  It says that a primordial object with mass \(10^{12}M_\odot\), if such
an object exists, would have to be associated with very early physics rather
than with ordinary late-time galactic accretion.

This distinction leaves open, but does not establish, a genuinely primordial
formation channel.  The minimal capture calculation is a late-time plasma
calculation performed in a simplified environment; it is not a no-go theorem
for rare early-universe processes.  Very rare peaks, phase transitions,
non-Gaussian charge or baryon fluctuations, defect-like configurations, or
other nonequilibrium plasma effects could in principle generate compact
objects whose initial conditions are not reproduced by ordinary accretion.
Such possibilities remain speculative unless an explicit mechanism is given.
They are nevertheless logically distinct from the calculation below, whose
negative result applies only to the ordinary accretion-limited branch.\\
 The benchmark
astron is not produced by the minimal capture model analyzed below.  That model
generates an ordinary accretion-limited branch with charges many orders of
magnitude below \(Q_A\).  The benchmark branch must therefore be read as a
primordial charge-concentration hypothesis whose consistency conditions are then
tested.  In particular, a viable model would have to explain both
\begin{equation}
  Q_A\sim 4\times10^{32}\,{\rm C}
\end{equation}
and the subsequent survival of this charge in a highly conducting plasma.

There is a useful numerical way to see why the charge is both small and
dynamically dangerous.  The number of elementary charges corresponding to the
fiducial value is
\begin{equation}
N_Q=\frac{Q_A}{e}
\simeq
2.5\times10^{51}
\left(\frac{Q_A}{4\times10^{32}\,{\rm C}}\right).
\end{equation}
The number of baryon masses associated with \(10^{12}M_\odot\) is
\begin{equation}
N_b\sim\frac{M_A}{m_p}
\simeq
1.2\times10^{69}
\left(\frac{M_A}{10^{12}M_\odot}\right).
\end{equation}
Hence the fractional charge imbalance is only
\begin{equation}
\epsilon_Q\equiv \frac{N_Q}{N_b}
\simeq
2\times10^{-18}
\left(\frac{Q_A}{4\times10^{32}\,{\rm C}}\right)
\left(\frac{10^{12}M_\odot}{M_A}\right).
\label{eq:fractional_charge_imbalance}
\end{equation}
The difficulty is therefore not that the required imbalance is large as a
particle-number fraction.  The difficulty is that electromagnetism is so much
stronger than gravity that this tiny imbalance can still correspond to
\(\Xi= k_eQ_A^2/(GM_A^2)\) of order unity or larger.  A primordial formation
model must explain how such a small imbalance is generated and, more
importantly, how it avoids being erased by the high conductivity of the early
plasma.

The same distinction also clarifies the cosmological question.  The original
phenomenological motivation for the large charge was the possibility that
long-range Coulomb repulsion among a sparse population of massive objects could
affect cosmic expansion.  However, a primordial origin for the objects does not
by itself imply late-time acceleration.  In a homogeneous FLRW reduction, the
interaction energy redshifts as \(a^{-4}\), as shown below, and therefore cannot
become an asymptotic cosmological constant.  Any acceleration associated with
astrons would have to arise from additional physics, such as a genuinely
inhomogeneous Einstein--Maxwell averaging effect.  This paper therefore treats
the benchmark as a consistency problem, not as a completed cosmological model.

The strategy of the paper is as follows.  Section~2 develops a minimal capture
model and shows that ordinary accretion produces only a much smaller
charge branch.  Section~3 compares this ordinary branch with the
phenomenological \(Q\propto M^2\) extrapolation.  Sections~4 and 5 discuss
charge persistence and screening, emphasizing that failure of the linear Debye
approximation is not proof of survival but a signal that nonlinear plasma
transport must be solved.  Section~6 adds the cosmological abundance and
discreteness constraints that any population of \(10^{12}M_\odot\) objects must
satisfy.  The later sections summarize the relevant compact-object geometry,
lensing and cosmological averaging issues, while the appendices are restricted
to short supporting formulae.

The resulting picture is deliberately conservative.  We do not claim to have
constructed a complete formation mechanism for the fiducial astron.  Rather, we
identify the quantitative obstacles that such a mechanism must overcome: the
charge-generation gap, nonlinear neutralization, and the large-scale-structure
constraints on a sparse population of ultra-massive compact objects.

\section{Charge Generation in a Minimal Capture Model}

We consider a collapsing overdensity of total mass $M$ and characteristic radius
$R(t)$ embedded in an ionized plasma of number density $n$ and temperature $T$.
The plasma is assumed to consist of electrons and protons with equal background
densities, so that the system is initially electrically neutral up to microscopic
fluctuations. The purpose of this section is modest: we use a minimal capture model
to estimate the timescale on which an incipient charge asymmetry can evolve and to
identify its parametric dependence.

The motivation for considering such a mechanism is not purely formal. In more standard
astrophysical settings, small equilibrium charges associated with proton-electron
mass asymmetry, as well as induced charges generated by rotating compact objects
immersed in magnetic fields, have been discussed extensively \cite{ZajacekTursunov2019,ZajacekEtAl2018}.

The motion of a particle of species $i$ with mass $m_i$ and charge $q_i$ in the field of a central object of mass $M$ and charge $Q$ is governed by the potential energy
\begin{equation}
U_i(r) = -\frac{GM m_i}{r} + \frac{k_e Q q_i}{r}.
\end{equation}
For the capture calculation it is convenient to divide by \(m_i\) and use the
potential per unit mass,
\begin{equation}
\phi_i(r) = -\frac{GM}{r} + \frac{k_e Q q_i}{m_i r}.
\end{equation}
The dynamics is most conveniently described in terms of the effective potential,
\begin{equation}
V_{\mathrm{eff}}(r) = \phi_i(r) + \frac{L^2}{2r^2},
\end{equation}
where $L$ is the angular momentum per unit mass. For a particle incoming from infinity with velocity $v_i$ and impact parameter $b_i$, one has
\begin{equation}
E = \frac{1}{2}v_i^2, \qquad L = b_i v_i.
\end{equation}

Capture occurs if the particle can reach the surface $r=R$, which requires that the radial kinetic energy vanishes at that point,
\begin{equation}
\frac{1}{2}v_i^2 = \phi_i(R) + \frac{b_i^2 v_i^2}{2R^2}.
\end{equation}
Solving for the critical impact parameter yields
\begin{equation}
b_i^2 = R^2\left(1 - \frac{2\phi_i(R)}{v_i^2}\right).
\end{equation}
The corresponding capture cross section is
\begin{equation}
\sigma_i = \pi b_i^2 = \pi R^2\left(1 - \frac{2\phi_i(R)}{v_i^2}\right).
\end{equation}

{\begin{reftwo}
This expression already shows the limitation of the focusing estimate.  If the
effective potential at the surface happens to satisfy \(\phi_i(R)=0\), the
geometrical interception cross section \(\sigma_i=\pi R^2\) remains.  Thus a
cancellation of the gravitational and electrostatic focusing terms does not by
itself mean that capture vanishes; it only marks the point at which the
strong-focusing approximation ceases to determine the capture rate.  The
force-balance scales below should be read in this restricted sense.
\end{reftwo}}

In the regime relevant for gravitational collapse, the focusing term dominates,
\begin{equation}
\frac{2|\phi_i(R)|}{v_i^2} \gg 1,
\end{equation}
and the cross section simplifies to
\begin{equation}
\sigma_i \simeq \frac{2\pi R}{v_i^2}
\left(GM-\frac{k_e Q q_i}{m_i}\right).
\end{equation}

Assuming a thermal distribution,
\begin{equation}
v_i^2 = \frac{k_B T}{m_i},
\end{equation}
one obtains
\begin{equation}
\sigma_i = \frac{2\pi R}{k_B T}(GM m_i - k_e Q q_i).
\end{equation}

The accretion rate is
\begin{equation}
\Gamma_i = n \sigma_i v_i
=
\frac{2\pi R n}{\sqrt{k_B T}}
\left(GM m_i^{1/2} - k_e Q q_i m_i^{-1/2}\right).
\end{equation}

The evolution of the net charge follows from the difference between electron and proton accretion,
\begin{equation}
\frac{dQ}{dt} = -e(\Gamma_e - \Gamma_p).
\end{equation}
Substituting the explicit expressions yields
\begin{equation}
\frac{dQ}{dt}
=
-\frac{2\pi R n e}{\sqrt{k_B T}}
\left[
GM(m_e^{1/2}-m_p^{1/2})
+k_eQe(m_e^{-1/2}+m_p^{-1/2})
\right].
\end{equation}

This equation takes the linear form
\begin{equation}
\frac{dQ}{dt} = C_0 + C_1 Q,
\label{eq:charge_linear}
\end{equation}
with
\begin{equation}
C_0 = -\frac{2\pi R n e GM}{\sqrt{k_B T}}(m_e^{1/2} - m_p^{1/2}), \qquad
C_1 = -\frac{2\pi R n k_e e^2}{\sqrt{k_B T}}(m_e^{-1/2} + m_p^{-1/2}).
\end{equation}

The term $C_0$ encodes the asymmetry between electron and proton capture in the
minimal model, while $C_1$ represents electric feedback. Since $m_p \gg m_e$, one
has $m_e^{1/2} - m_p^{1/2} < 0$, so that $C_0 > 0$ and $C_1 < 0$ within the conventions
used here.

For fixed local plasma parameters \((R,n,T)\), this linear equation can be solved
exactly for an arbitrary initial charge \(Q(0)=Q_0\). One finds
\begin{equation}
Q(t)=Q_{\mathrm{eq}}+\left(Q_0-Q_{\mathrm{eq}}\right)e^{C_1 t},
\qquad
Q_{\mathrm{eq}}=-\frac{C_0}{C_1}.
\label{eq:charge_solution_general}
\end{equation}
Since \(C_1<0\), the solution tends asymptotically toward the fixed point
\(Q_{\mathrm{eq}}\). If one starts from a neutral object, \(Q_0=0\), then
\begin{equation}
Q(t)=Q_{\mathrm{eq}}\left(1-e^{-t/\tau}\right),
\qquad
\tau=\frac{1}{|C_1|}
=
\frac{\sqrt{k_B T}}{2\pi R n k_e e^2 (m_e^{-1/2}+m_p^{-1/2})}.
\label{eq:tau_response}
\end{equation}
At early times \(Q\ll Q_{\mathrm{eq}}\), the growth is approximately linear,
\begin{equation}
Q(t)\simeq C_0 t,
\end{equation}
while at later times the electric feedback proportional to \(C_1Q\) slows the
approach to the fixed point. Thus \(\tau\) is simply the characteristic charging
response time in the linearized model. It measures how quickly the charge moves
toward \(Q_{\mathrm{eq}}\), but it does not by itself determine the exact
equilibration time or the final nonlinear saturation value.

The charging timescale is controlled primarily by the product $Rn$,
with only a square-root dependence on temperature. Numerically, one may rewrite
it as
\begin{equation}
\tau \approx 2.39\,
\frac{\sqrt{T/\mathrm{K}}}{R(\mathrm{m})\,n(\mathrm{m}^{-3})}\,\mathrm{s}.
\label{taunum_env}
\end{equation}
It is therefore essential to state clearly which plasma environment is being used
when quoting any benchmark value for $\tau$: the charging timescale is not a
fixed number. Depending on the plasma environment and on the effective capture
radius, the formal response time can range from seconds in an extremely dilute
benchmark to microscopic values in dense formation regions (see Appendix A). It
is therefore more informative to speak of a \emph{range} of charging timescales
than of a single benchmark. The appropriate value of $\tau$ must always be tied
to a specific choice of $(R,n,T)$ appropriate to the physical stage of the astron
formation process under consideration.

The physical interpretation is straightforward. The charging time decreases when the
local plasma density $n$ increases, because a denser environment supplies more particles
to the capture process. It also decreases when the effective capture radius $R$ is
larger, since a larger interaction region enhances the accretion rate. By contrast,
increasing the temperature tends to lengthen the charging time, but only through
a square-root dependence, so this effect is comparatively mild.

To compare this response time with the collapse time in a meaningful way, it is
useful to evaluate both quantities in the same local environment. The free-fall time of a self-gravitating gas of density $\rho$ is
\begin{equation}
t_{\mathrm{ff}}=\sqrt{\frac{3\pi}{32G\rho}}.
\label{eq:tff_env_comp}
\end{equation}
(see appendix A).
For an ionized hydrogen plasma one may take
\begin{equation}
\rho \simeq n m_p,
\end{equation}
so that
\begin{equation}
t_{\mathrm{ff}}
=
\sqrt{\frac{3\pi}{32Gm_p n}}
\approx
1.62\times10^{18}\left(\frac{n}{\mathrm{m}^{-3}}\right)^{-1/2}\mathrm{s}.
\label{eq:tff_n_env_comp}
\end{equation}

In order to illustrate the range of possibilities, we use a deliberately compact
reference capture radius, normalized to the gravitational radius of a fiducial
astron of mass \(M\sim 10^{12}M_\odot\),
\begin{equation}
R \sim R_g = \frac{GM}{c^2}
\simeq 1.48\times10^{15}\,\mathrm{m}
\simeq 4.8\times10^{-2}\,\mathrm{pc}
\simeq 9.9\times10^{3}\,\mathrm{AU}.
\end{equation}
This is a compactness benchmark, not a statement that the physical surface must
sit at \(R_g\); for a neutral object the black-hole threshold would be set by
\(2R_g\), while for a charged object it is set by the RN horizon condition.
The corresponding values of $\tau$ and $t_{\mathrm{ff}}$ then depend entirely on
the local plasma environment.

For the present-day baryon density, the Particle Data
Group quotes $\Omega_b h^2\simeq 0.02237$, corresponding to a mean baryon density
today of order $n_{b,0}\sim 2.5\times10^{-7}\,\mathrm{cm}^{-3}$ \cite{PDG2024,PDG2025Astro}.
\begin{pubrev}
The conversion and its standard FLRW interpretation are consistent with the
cosmological conventions summarized in Refs.~\cite{Weinberg2008,DodelsonSchmidt2020}.
\end{pubrev}

By contrast, much denser environments arise in direct-collapse and supermassive-seed
formation scenarios. High-velocity protogalactic collisions can produce shocked gas
with $T\sim 10^6\,\mathrm{K}$ and $n\gtrsim 10^4\,\mathrm{cm}^{-3}$ \cite{InayoshiVisbalKashiyama2015};
cosmological radiation-hydrodynamics simulations of direct-collapse clouds follow the
gas at least up to central densities of order $10^8\,\mathrm{cm}^{-3}$ \cite{ChonHosokawaYoshida2018};
and radiative-transfer calculations of the optically thick inner collapse report densities
as large as $\rho\sim 10^{-6}\,\mathrm{g\,cm}^{-3}$, corresponding to number densities
well above $10^{20}\,\mathrm{m}^{-3}$ for ionized hydrogen \cite{LuoEtAl2018}.

Using these environments as order-of-magnitude benchmarks, the corrected capture
estimate gives a formal response time shorter than the corresponding free-fall
time even for very dilute benchmark densities, and microscopic values in dense
formation-stage environments. This should not be overinterpreted as proof that a
diffuse late-time IGM can build the large phenomenological charge: in such a
medium the limiting issues are the finite charge reservoir, transport, and
screening, not the local linear response coefficient. The comparison
\begin{equation}
\tau \ll t_{\mathrm{ff}}
\end{equation}
should however be interpreted with care. In the derivation of \(\tau\), the local
background quantities \(R\), \(n\), and \(T\) are treated as fixed, whereas in an
actual collapse they evolve together with the infall and with the charge itself.
Moreover, as \(Q\) grows, the Coulomb force feeds back on the incoming trajectories
and can slow or shut off further capture. The present comparison with \(t_{\mathrm{ff}}\)
is therefore not a self-consistent solution of the coupled collapse-plus-charging
problem. Its meaning is more limited but still useful: within the assumptions of
the local capture model, the charge sector adjusts much faster than the bulk
hydrodynamic background changes. A fully consistent treatment would require solving
simultaneously for the collapse flow, the evolving density and temperature profiles,
and the charge-dependent accretion dynamics; we defer that final-radius problem to
a separate analysis.

The capture asymmetry and the electric
feedback define a preferred charge scale and a short response time, while the final
sign and nonlinear endpoint depend on physics that is not fully encoded in the linearized
equation alone. This motivates the discussion of saturation in the next section.

\section{Nonlinear Growth, Saturation and a Scaling Extrapolation}

The previous section identifies a short characteristic response time for the charge
sector. Once the charge becomes appreciable, however, the electric contribution is
no longer a perturbation, and nonlinear effects must control the subsequent evolution.
We now analyze the mechanism that limits further growth.

\begin{pubrev}
The full evolution equation derived in Section 2 is
Eq.~\eqref{eq:charge_linear}. It is valid only as long as the capture cross
sections remain positive for both
\end{pubrev}
species. As the magnitude of the charge increases, the Coulomb term modifies the
effective potential differently for electrons and protons, leading to a progressive
suppression of one channel and enhancement of the other. It is therefore important
to keep the sign convention explicit. We take
\begin{equation}
q_p=+e,\qquad q_e=-e,
\end{equation}
so that a positive source charge attracts electrons and repels protons.

The neutral initial condition of the minimal model naturally moves toward the
positive-charge branch, because proton capture is gravitationally focused more
efficiently than electron capture. For \(Q>0\), the surface potential energies are
\begin{equation}
U_e(R) = -\frac{GM m_e}{R} - \frac{k_e Qe}{R}, \qquad
U_p(R) = -\frac{GM m_p}{R} + \frac{k_e Qe}{R}.
\end{equation}
As \(Q\) grows, electrons are attracted more strongly, while proton capture is
progressively suppressed. {\begin{reftwo}The gravitational focusing of the
proton channel is cancelled when\end{reftwo}}
\begin{equation}
GM m_p = k_e Q e.
\end{equation}
The corresponding charge scale is
\begin{equation}
Q_p=\frac{GMm_p}{k_ee}.
\end{equation}
This is not a runaway point: if \(Q\) becomes too large, electron capture reduces
the positive charge, while if \(Q\) is too small the proton excess in the capture
rate increases it. In the linear focusing approximation the exact fixed point is
\begin{equation}
\label{eq:current_balance_charge}
Q_{\mathrm{eq}}
=
\frac{GM}{k_e e}
\frac{m_p^{1/2}-m_e^{1/2}}{m_p^{-1/2}+m_e^{-1/2}}
\simeq
\frac{GM\sqrt{m_pm_e}}{k_e e},
\end{equation}
and perturbations obey
\begin{equation}
\frac{d\,\delta Q}{dt}=C_1\delta Q,
\qquad C_1<0,
\end{equation}
so the fixed point is linearly stable.

One can also define the opposite, negative-charge branch. For \(Q=-|Q|<0\),
\begin{equation}
U_e(R) = -\frac{GM m_e}{R} + \frac{k_e |Q|e}{R}, \qquad
U_p(R) = -\frac{GM m_p}{R} - \frac{k_e |Q|e}{R}.
\end{equation}
In that case electrons are repelled and protons are attracted. The electron channel
{\begin{reftwo}loses its gravitational focusing when\end{reftwo}}
\begin{equation}
GM m_e = k_e |Q| e,
\end{equation}
giving the lower charge scale
\begin{equation}
|Q_e|=\frac{GMm_e}{k_ee}.
\end{equation}
{\begin{reftwo}
This negative branch is therefore a force-balance scale for an imposed
negative source charge.  It is not a second stable fixed point of the same
linear rate equation that evolves a neutral object toward the positive
current-balance value.
\end{reftwo}}
Thus the corresponding focusing-cancellation scales may be written compactly as
\begin{equation}
\label{micro}
|Q|_{\mathrm{foc}}\sim \frac{G M m_*}{k_e e},
\end{equation}
where \(m_*=m_p\) for the natural positive branch generated by proton-dominated
capture, and \(m_*=m_e\) for an imposed negative branch. These ordinary
{\begin{refthree}scales mark cancellation of the gravitational-focusing term
for the indicated species.  They are not dynamically selected equilibrium
charges.  The dynamically selected charge within the linear capture model is
instead the current-balance fixed point \(Q_{\rm eq}\) in
Eq.~\eqref{eq:current_balance_charge}, where the
electron and proton capture currents are equal.  Both kinds of scale should be
distinguished from the much larger phenomenological Frampton-type extrapolation
discussed below.\end{refthree}}

A complementary way to estimate the onset of saturation is to compare the outward
stress generated by the electric field with the inward ram pressure of the infalling
plasma. The electric field acts mechanically on the surface through the Maxwell
stress tensor, so its effect can be expressed as an effective electrostatic pressure.

\begin{equation}
P_E = n_i T_{ij} n_j = T_{rr},
\qquad
T_{ij}=\epsilon_0\left(E_iE_j-\frac{1}{2}\delta_{ij}E^2\right).
\end{equation}
Here $T_{ij}$ is the electrostatic Maxwell stress tensor in SI units, and the pressure is
the normal component of the field stress. For a spherical charged surface the normal
direction is radial, so the relevant component is $T_{rr}$.

\begin{equation}
P_E = T_{rr}=\frac{\epsilon_0 E_r^2}{2}
\sim
\frac{\epsilon_0}{2}\left(\frac{k_e Q}{R^2}\right)^2
= \frac{k_eQ^2}{8\pi R^4}.
\end{equation}
In the last step we used the radial electric field at the surface of a spherically
symmetric charge distribution, $E_r(R)\sim k_eQ/R^2$, which shows that the field produces
an outward electrostatic pressure scaling as \(k_eQ^2/R^4\).

{\begin{refthree}
The microscopic capture calculation therefore contains two conceptually
different statements.  Its rate equation drives a neutral object toward the
stable current-balance value \(Q_{\rm eq}\).  Separately, the force-balance
estimates identify charges at which gravitational focusing of one species is
cancelled.  The latter diagnose the breakdown of the strong-focusing
approximation; they do not establish a second dynamical equilibrium or complete
cessation of geometric capture.
\end{refthree}}

The same saturation condition can be recovered in a complementary macroscopic
language by comparing the outward electrostatic stress with the inward ram pressure
of the accreting plasma. If the electric field becomes strong enough that its
surface stress matches the momentum flux of the inflow, further accretion must
cease. The inward ram pressure of the infalling plasma is
\begin{equation}
P_{\mathrm{ram}} \sim \rho v^2,
\end{equation}
where $\rho$ is the mass density of the inflow and $v$ its characteristic speed.
The condition for continued accretion is therefore
\begin{equation}
P_{\mathrm{ram}} \gtrsim P_E.
\end{equation}
Equivalently, saturation is reached when
\begin{equation}
P_E \sim P_{\mathrm{ram}},
\end{equation}
that is,
\begin{equation}
\frac{k_eQ^2}{8\pi R^4} \sim \rho v^2.
\end{equation}
Solving for the charge gives the corresponding equilibrium scale
\begin{equation}
Q_{\mathrm{sat}} \sim \sqrt{\frac{8\pi \rho v^2}{k_e}}\,R^2.
\end{equation}
Beyond this point the electric stress dominates the inflow stress, so incoming
charged particles are no longer able to accrete efficiently.
The pressure-balance estimate should not be expected to reproduce exactly the same
numerical coefficient as the microscopic force-balance criterion, since it is a
coarse-grained argument. However, it does recover the same parametric saturation
scale. Let $m_*$ denote the mass of the particle species whose accretion is being
suppressed at saturation. To compare the microscopic condition \eqref{micro} with the pressure picture, write $|Q|=Ne$, so that the number of captured
carriers is $N\sim |Q|/e$. If these occupy a region of size $R$, then
\begin{equation}
n_* \sim \frac{3|Q|}{4\pi eR^3},
\qquad
\rho_* \sim m_* n_* \sim \frac{3m_*|Q|}{4\pi eR^3}.
\end{equation}
Taking the inflow speed to be of order the free-fall speed, $v^2\sim GM/R$, the ram
pressure becomes
\begin{equation}
P_{\mathrm{ram}} \sim \rho_* v^2 \sim \frac{3GMm_*|Q|}{4\pi eR^4}.
\end{equation}
Equating this with the electrostatic pressure
\begin{equation}
P_E \sim \frac{k_e Q^2}{8\pi R^4}
\end{equation}
yields
\begin{equation}
|Q|_{\mathrm{sat}} \sim \frac{6GMm_*}{k_e e}.
\end{equation}
Thus the pressure formalism reproduces the same saturation scale up to a factor of
order unity. The agreement is therefore parametric rather than exact, as expected
for two different approximate descriptions of the same physical mechanism.
For an astron of mass
\begin{equation}
M_A = 10^{12} M_\odot \simeq 1.99 \times 10^{42}\ \mathrm{kg},
\label{eq:fiducial_astron_mass}
\end{equation}
the ordinary accretion-limited saturation charge is obtained from
\begin{equation}
|Q|_{\mathrm{sat}} \sim \frac{G M_A m_*}{k_e e},
\end{equation}
where \(m_*\) is the mass of the particle species whose accretion is being
suppressed.

Using
\begin{equation}
G = 6.6743\times 10^{-11}\ \mathrm{m^3\,kg^{-1}\,s^{-2}},\qquad
k_e = 8.9876\times 10^{9}\ \mathrm{N\,m^2\,C^{-2}},
\end{equation}
\begin{equation}
e = 1.6022\times 10^{-19}\ \mathrm{C},\qquad
m_e = 9.1094\times 10^{-31}\ \mathrm{kg},\qquad
m_p = 1.6726\times 10^{-27}\ \mathrm{kg},
\end{equation}
the sign convention must be kept explicit. We define
\(Q=e(N_p-N_e)\), so a positive object repels protons and attracts electrons,
whereas a negative object repels electrons and attracts protons. Thus the
electron-limited branch is a negative-charge branch, while the proton-limited
branch is a positive-charge branch. With signed charges one finds
\begin{equation}
Q_{\mathrm{sat}}^{(e)}
=
-\frac{G M_A m_e}{k_e e}
=
-\frac{(6.6743\times10^{-11})(1.99\times10^{42})(9.1094\times10^{-31})}
     {(8.9876\times10^{9})(1.6022\times10^{-19})}
\simeq -8.4\times10^{10}\ \mathrm{C},
\end{equation}
while for proton-limited saturation
\begin{equation}
Q_{\mathrm{sat}}^{(p)}
=
\frac{G M_A m_p}{k_e e}
=
\frac{(6.6743\times10^{-11})(1.99\times10^{42})(1.6726\times10^{-27})}
     {(8.9876\times10^{9})(1.6022\times10^{-19})}
\simeq 1.5\times10^{14}\ \mathrm{C}.
\end{equation}

{\begin{reftwo}
It is useful to distinguish these force-balance scales from the linear
rate-balance value. The force-balance estimates above ask when the Coulomb force
on one species is large enough to cancel the gravitational focusing contribution
for that species. The
linear rate-balance value asks a different question: for a fixed background,
what charge makes the proton and electron capture rates equal? Starting from
\(Q=0\), the strong-focusing rate is slightly larger for protons than for
electrons, so the object first charges positively. A positive \(Q\) then
suppresses proton capture and enhances electron capture until
\end{reftwo}}
\begin{equation}
\Gamma_p(Q_{\rm eq}^{(+)})=\Gamma_e(Q_{\rm eq}^{(+)}).
\end{equation}
Using the strong-focusing rates gives
\begin{equation}
Q_{\mathrm{eq}}^{(+)}
=
\frac{GM_A}{k_e e}
\frac{\sqrt{m_p}-\sqrt{m_e}}
{m_e^{-1/2}+m_p^{-1/2}}
=
\frac{GM_A\sqrt{m_pm_e}}{k_e e}
\frac{\sqrt{m_p}-\sqrt{m_e}}
{\sqrt{m_p}+\sqrt{m_e}} .
\end{equation}
Since \(m_p\gg m_e\), the last factor is close to one. Therefore
\begin{equation}
Q_{\mathrm{eq}}^{(+)}
\simeq
\frac{G M_A\sqrt{m_pm_e}}{k_e e}
\simeq 3.6\times10^{12}\,\mathrm C .
\end{equation}
This value lies between the electron- and proton-limited force-balance
magnitudes because \(\sqrt{m_pm_e}\) is the geometric mean of \(m_e\) and
\(m_p\). It is the equilibrium charge of the linear capture equation, not a
separate condition for complete shutdown of an accretion channel.

Thus, within the ordinary accretion picture, the two focusing-cancellation
scales for an astron of mass \(10^{12}M_\odot\) span the range
\begin{equation}
10^{11}\ \mathrm{C}\ \lesssim\ |Q|_{\mathrm{foc}}\ \lesssim\ 10^{14}\ \mathrm{C},
\end{equation}
depending on the species.  The actual fixed point of the linear capture
equation is the distinct current-balance value
\(Q_{\rm eq}^{(+)}\simeq3.6\times10^{12}\,\mathrm C\).  In the plots below the
force-balance curves are shown as magnitudes; the physical sign is fixed by
which species is being repelled.

This result is much smaller than the fiducial phenomenological charge
\begin{equation}
Q_A\sim4\times10^{32}\,\mathrm C .
\end{equation}
The gap is
\begin{equation}
\frac{Q_A}{|Q|_{\rm foc}}
\sim 10^{18}\hbox{--}10^{21}.
\end{equation}
This is the most important conclusion of the minimal capture calculation:
ordinary accretion-driven charge separation does not produce the fiducial
astron.  The large-charge branch must be supplied by a different primordial
charge-concentration mechanism, or else it should be rejected.  The role of the
capture model in this paper is therefore diagnostic.  It establishes the scale
of ordinary charging and exposes the size of the missing mechanism.

\begin{figure}[t]
\centering
\includegraphics[width=0.64\textwidth]{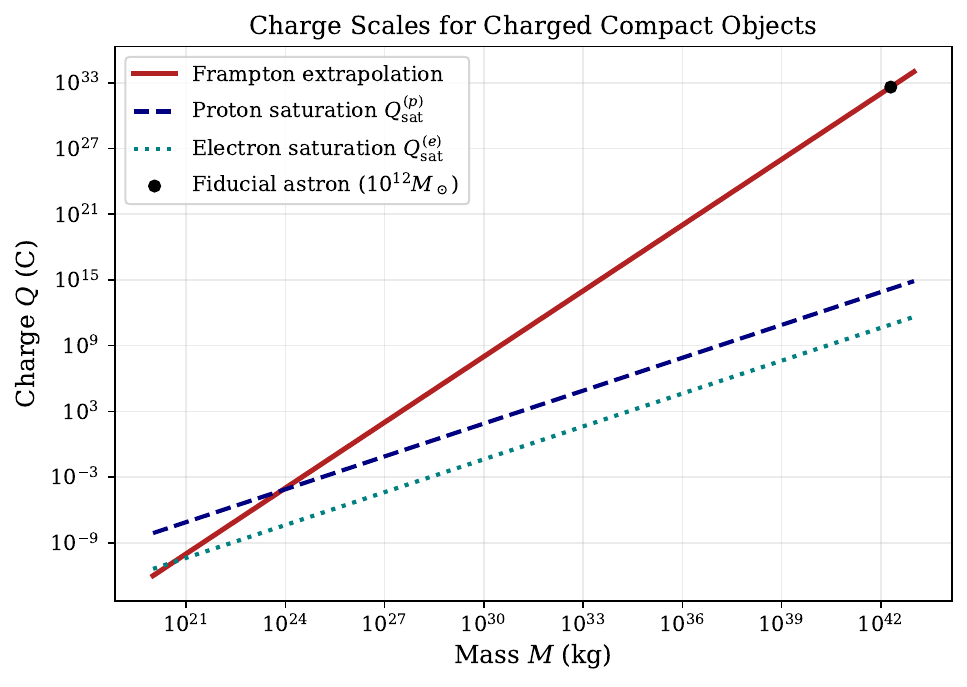}
\caption{Numerical scan of the charge scales relevant to the astron scenario.
The red curve shows the Frampton mass--charge extrapolation \(Q=10^{-52}M^2\),
while the blue and teal curves show the ordinary proton- and electron-limited
saturation estimates \(Q_{\rm sat}\sim G M m_*/(k_e e)\). The fiducial
\(10^{12}M_\odot\) astron is marked explicitly. The plot makes clear that the
Frampton branch occupies a parametrically different regime from ordinary
accretion-limited charging.}
\label{fig:charge_scales}
\end{figure}

It is useful to summarize these scales in a single numerical scan. Although the
plots that follow are not dynamical simulations, they provide a compact global
view of the analytic relations derived in this section. Figure~\ref{fig:charge_scales}
compares the ordinary proton- and electron-limited saturation charges with the
Frampton extrapolation \(Q\propto M^2\). The qualitative separation is already
clear at the level of the raw charge: the extrapolated branch grows so rapidly
with \(M\) that by \(M\sim10^{12}M_\odot\) it lies many orders of magnitude
above the ordinary accretion-limited values.

\subsection{The $Q/M$ extrapolation}

A different estimate was proposed in Ref.~\cite{Frampton2022EAU}, where the
charge is not derived from a local accretion-saturation condition but from an
assumed mass--charge scaling for extremely massive primordial black holes.
The basic idea is to ask when the electric repulsion between two identical
objects can become comparable to, or larger than, their mutual gravitational
attraction.

For two objects of mass \(M\) and charge \(Q\), the forces scale as
\(F_G \sim GM^2/r^2\) and \(F_E \sim k_e Q^2/r^2\), so that the relevant ratio is
\(R \equiv F_G/F_E \sim GM^2/(k_e Q^2)\). If \(R>1\), gravity dominates, whereas
if \(R<1\), Coulomb repulsion dominates.

Ref.~\cite{Frampton2022EAU} then adopts
two benchmark values for the charge-to-mass ratio from the PBH-charging analysis
of Ref.~\cite{ArayaEtAl2022} and extends them through a log-linear ansatz,
which in practice implies \(Q/M \propto M\) and hence \(Q \propto M^2\). Applied
to the fiducial astron mass in Eq.~\eqref{eq:fiducial_astron_mass}, this
extrapolation gives the large-charge benchmark
\begin{equation}
Q_A \simeq 4\times10^{32}\,\mathrm{C}.
\label{eq:fiducial_large_charge}
\end{equation}
This is the scale at which electromagnetic repulsion is argued to become
cosmologically significant. We have reported in Appendix
\ref{app:qm_extrapolation} details of the derivation.

Although the extrapolation used in Ref.~\cite{Frampton2022EAU} is admittedly
hypothetical, the regime to which it is applied is itself far beyond any
empirically controlled theory of primordial ultra-massive object formation.
In that sense, the proposal should be read as an exploratory phenomenological
extension.

The broader early-universe context of this assumption was discussed in
Section~\ref{subsec:primordial_status}.  The extrapolation should be treated as
a phenomenological working hypothesis whose physical consequences must be tested
directly; it is not justified by observations of early luminous structure alone.

In the next sections we examine the physical consequences of this extrapolation,
with particular attention to the issues of Debye screening and to the spacetime
interpretation of such highly charged objects in the super-extremal, naked-singularity
regime.

In the analysis that follows we shall assume that the characteristic inter-astron
separation is of order megaparsecs. This is the physically relevant regime if
astrons are to play a role in cosmological acceleration, since both the strength
of their collective Coulomb interaction and the possible impact of Debye
screening must be evaluated on scales comparable to the mean separation of the
population.

\begin{figure}[t]
\centering
\includegraphics[width=0.64\textwidth]{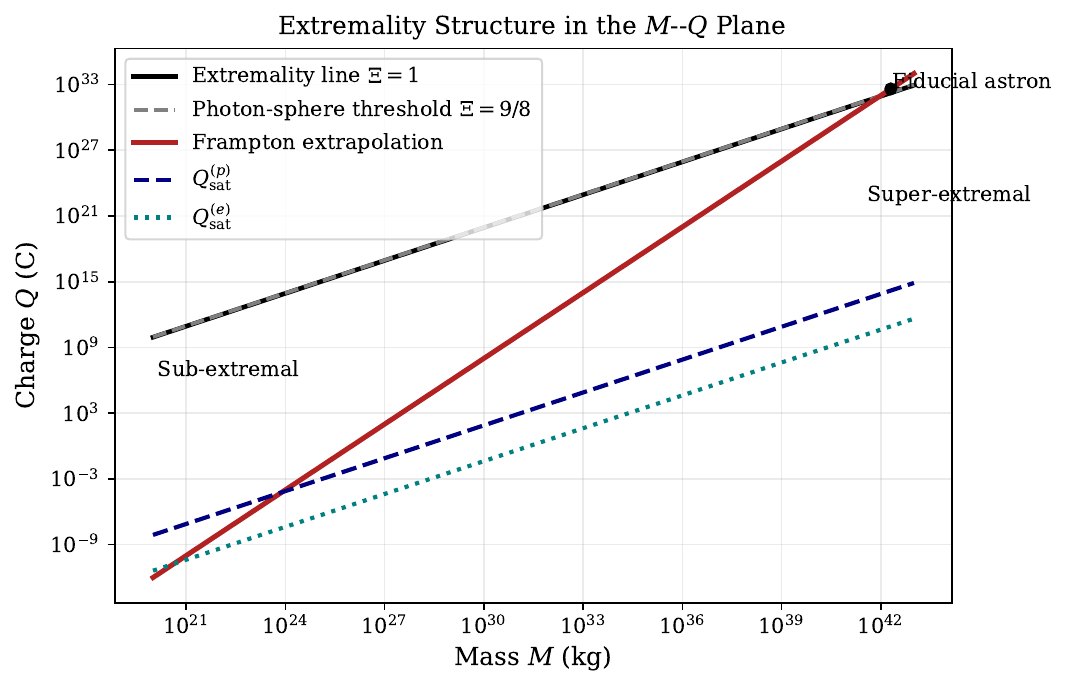}
\caption{Extremality structure in the \((M,Q)\) plane. The black line marks the
Reissner--Nordstr\"om extremality bound \(\Xi=1\), the gray dashed line marks
the photon-sphere threshold \(\Xi=9/8\), and the colored curves show the same
charge prescriptions as in Fig.~\ref{fig:charge_scales}. The fiducial astron
point lies deep in the super-extremal region, while the ordinary saturation
branches remain many orders of magnitude below extremality.}
\label{fig:extremality_phase}
\end{figure}

The same numerical scan can be represented directly in the \((M,Q)\) plane,
where the geometric significance of the extrapolated charge becomes more
transparent. Figure~\ref{fig:extremality_phase} shows the Frampton branch
together with the Reissner--Nordstr\"om extremality line \(\Xi=1\) and the
photon-sphere threshold \(\Xi=9/8\). The extrapolated branch crosses into the
super-extremal regime at masses of order a few \(10^{11}M_\odot\), whereas the
ordinary saturation curves remain extremely far below both thresholds over the
entire range shown. This anticipates the main conclusion of Section~6: the
large-charge astron proposal and the ordinary accretion-saturation estimate do
not correspond to nearby points in parameter space, but to qualitatively
different Einstein--Maxwell regimes.

\section{Stability of the Charge}

The persistence of a macroscopic electric charge over cosmological timescales
requires the absence of efficient discharge or neutralization mechanisms. This
issue involves three conceptually distinct effects: plasma neutralization by the
surrounding medium, continued accretion of opposite charges, and quantum vacuum
discharge through pair production.

From the astrophysical point of view, the possibility that black holes may carry
small but nonvanishing electric charges has been revisited in recent years, especially
because even modest charges can affect the motion of charged particles and modify
radiative signatures \cite{ZajacekTursunov2019,ZajacekEtAl2018}. Our purpose here is
to examine whether an analogous conclusion can hold, at least parametrically, for
the much more highly charged compact objects envisioned in the astron scenario.

We first consider neutralization by the intergalactic medium (IGM). For the
fiducial large-charge branch in Eq.~\eqref{eq:fiducial_large_charge},
the total number of elementary charges required to compensate the astron charge is
\begin{equation}
\label{comp}
N_Q = \frac{Q_A}{e}
\simeq
\frac{4\times10^{32}}{1.602\times10^{-19}}
\simeq 2.5\times10^{51}.
\end{equation}
As a representative present-day benchmark, we take the intergalactic electron density
to be of order
\begin{equation}
n_e \sim 0.2\ \mathrm{m^{-3}},
\label{eq:igm_ne_benchmark}
\end{equation}
consistent with the mean cosmological baryon density and standard descriptions of the
low-density ionized IGM
\begin{pubrev}\cite{PDGAstro,CfAIGM,COSMOSIGM,McQuinn2016,DaveEtAl2001,TumlinsonPeeplesWerk2017}\end{pubrev}.
The number of available
free charges within a sphere of radius \(R\) is
\begin{equation}
N_{\mathrm{IGM}}(R)=n_e\frac{4\pi}{3}R^3.
\label{eq:nigm_sphere}
\end{equation}
For \(R=1\,\mathrm{pc}=3.086\times10^{16}\,\mathrm m\), this gives
\begin{equation}
N_{\mathrm{IGM}}(1\,\mathrm{pc})
\simeq
(0.2)\frac{4\pi}{3}(3.086\times10^{16})^3
\simeq 2.5\times10^{49}.
\end{equation}
Thus a one-parsec sphere falls short of the fiducial compensation charge by roughly
two orders of magnitude. Complete neutralization would require an extended charge
reservoir, with a characteristic radius of order several parsecs for this mean
density. This counting argument does not replace a plasma-screening calculation,
but it shows that local neutralization is not a compact near-surface process.

The remaining question is whether a large charge can subsequently be removed by quantum
vacuum effects. The relevant process is the Schwinger mechanism, namely the spontaneous
creation of electron--positron pairs when the electric field becomes so intense that the
vacuum itself is unstable. The corresponding critical field is
\begin{equation}
E_{\mathrm{crit}}=\frac{m_e^2 c^3}{e\hbar}
\simeq 1.3\times10^{18}\ \mathrm{V/m}.
\end{equation}
If the electric field near the charged object were to approach this value, pairs would
be produced, with charges of opposite sign falling toward the source and thereby tending
to reduce its net charge.

For an astron the electric field outside the source is
\begin{equation}
E(r)=\frac{k_e Q_A}{r^2}.
\end{equation}
Evaluated at the reference compactness scale
\begin{equation}
r_A=\frac{G M_A}{c^2}\simeq1.5\times10^{15}\,\mathrm m.
\label{eq:astron_grav_scale}
\end{equation}
one finds
\begin{equation}
E(r_A)\sim 1.6\times10^{12}\ \mathrm{V/m},
\end{equation}
for the fiducial mass in Eq.~\eqref{eq:fiducial_astron_mass} and the charge in
Eq.~\eqref{eq:fiducial_large_charge}. Hence
\begin{equation}
\frac{E(r_A)}{E_{\mathrm{crit}}}\sim 1.2\times10^{-6},
\end{equation}
so the electric field at this compactness scale is many orders of magnitude below
the Schwinger threshold. {\begin{reftwo}For a regular finite-radius realization
with physical surface \(R_s\gtrsim r_A\), the exterior field at the surface is
therefore subcritical, and it decreases further as \(r^{-2}\).  In that case
the exterior Schwinger channel is strongly suppressed.  If instead \(R_s<r_A\),
or if one takes the exact super-extremal RN singularity literally, the relevant
criterion is not the reference compactness scale but the comparison of the
physical inner cutoff with \(r_{\rm crit}\) below.\end{reftwo}}

One may still ask whether the Schwinger threshold could be reached only at much smaller
radius. Solving
\begin{equation}
E(r_{\mathrm{crit}})=E_{\mathrm{crit}}
\end{equation}
gives
\begin{equation}
r_{\mathrm{crit}}=\sqrt{\frac{k_e Q_A}{E_{\mathrm{crit}}}}.
\end{equation}
{\begin{reftwo}
For the astron parameters considered here, this radius lies well below the
compactness scale.  This is sufficient to suppress pair production outside a
regular body whose surface lies at or above the compactness scale.  It is not,
by itself, a complete exclusion of the Schwinger channel for an exact
super-extremal RN singularity, because no horizon hides the region
\(r\lesssim r_{\rm crit}\).  In that singular realization, the critical-field
region requires either a specified inner cutoff, a UV completion, or a
semiclassical treatment of the near-source electrovac region.
\end{reftwo}}
For the benchmark astron charge in Eq.~\eqref{eq:fiducial_large_charge},
\(k_e=8.99\times10^9\,\mathrm{N\,m^2\,C^{-2}}\), and
\(E_{\mathrm{crit}}\simeq1.3\times10^{18}\,\mathrm{V/m}\), one finds
\begin{equation}
r_{\mathrm{crit}}
=
\sqrt{\frac{k_e Q_A}{E_{\mathrm{crit}}}}
=
\sqrt{\frac{(8.99\times10^9)(4\times10^{32})}{1.3\times10^{18}}}
\simeq 1.7\times10^{12}\,\mathrm m.
\end{equation}
For the fiducial astron mass in Eq.~\eqref{eq:fiducial_astron_mass}, the
reference compactness scale is the one given in
Eq.~\eqref{eq:astron_grav_scale}.
Hence
\begin{equation}
\frac{r_{\mathrm{crit}}}{r_A}\simeq 1.1\times10^{-3},
\end{equation}
so the Schwinger threshold would be reached only at radii roughly three orders
of magnitude smaller than the gravitational scale of the object.

{\begin{reftwo}
The Schwinger estimate should therefore be read as a constraint on one
discharge channel, not as a general proof of charge survival.  For a regular
astron with \(R_s\gg r_{\rm crit}\), the exterior field is subcritical
everywhere outside the material surface.  For an exact RN singularity, the
critical-field region is part of the formal exterior and cannot be dismissed
without an additional physical prescription.
\end{reftwo}}

This conclusion is limited.  {\begin{refthree}For the regular finite-radius
realization assumed in the numerical estimate, with
\(R_s\gg r_{\rm crit}\), the exterior field is subcritical and Schwinger
discharge is strongly suppressed outside the material surface.  This statement
does not apply to an exact super-extremal RN singularity: in that case the
formal exterior includes the critical-field region and the discharge problem
remains unresolved without an inner cutoff, ultraviolet completion, or
semiclassical near-source analysis.\end{refthree}}  The regular-object estimate
does not show that the charge survives nonlinear plasma neutralization.  A
large positive potential strongly attracts electrons, and in a conducting
medium the relevant question is whether electron inflow is prevented or slowed
by a real kinetic bottleneck.  Possible ingredients include angular momentum,
magnetic fields, electrostatic trapping, depletion of the available reservoir,
collisionless phase-space constraints, or cosmological expansion.  None of
these is demonstrated by the minimal model.

Thus the status of charge persistence is an open consistency condition.  A
viable astron model must show that the nonlinear neutralization time is longer
than the relevant formation, dynamical, or cosmological time.  {\begin{refthree}
Suppression of the exterior Schwinger channel for the regular, subcritical
configuration removes one possible discharge mechanism for that configuration
only; it does not establish survival of the fiducial charge and does not settle
the singular RN case.\end{refthree}}

\section{Screening Constraints in the Intergalactic Medium}

A central assumption of the astron scenario is that the electric field generated by
a charged compact object can remain effective over distances comparable with the
mean inter-astron separation. Since astrons are embedded in an ionized cosmic
environment, this immediately raises the question of plasma screening. In an
ordinary quasi-neutral electron--ion plasma, a localized charge is not felt out to
arbitrarily large distances: mobile charges of the opposite sign are attracted,
while charges of the same sign are repelled, producing a polarization cloud that
reduces the effective field beyond a characteristic screening scale. The standard
language for this phenomenon is Debye screening. This should not be confused with
a gas of same-sign charges alone. If only positive charges are present, there is
no ordinary Debye screening of a positive source unless a compensating negative
component or neutralizing background is also introduced.

For the present problem, the main issue is not whether screening exists in some
form, but whether the linear Debye picture can be applied without modification
to an object as extreme as an astron. On the one hand, the intergalactic medium
is very dilute, which might suggest weak screening. On the other hand, the
charges postulated for astrons are so large that the electrostatic potential near
the source is far from the weak-perturbation regime assumed in the usual linear
Debye--H\"uckel treatment. The standard formulae are therefore useful as a first
benchmark, but they cannot by themselves settle the question.

In this section we use the standard Debye result only as a benchmark; the
derivation is given in Appendix~\ref{debye}. For a quasi-neutral multicomponent
plasma in linear response,
\begin{equation}
\frac{1}{\lambda_D^2}
=
\sum_s\frac{n_{s0}q_s^2}{\epsilon_0 k_BT_s},
\label{eq:debye_multicomponent}
\end{equation}
where \(n_{s0}\), \(q_s\), and \(T_s\) are the unperturbed density, charge, and
temperature of species \(s\). A point charge \(Q\) then produces the screened
potential
\begin{equation}
\Phi(r)=\frac{Q}{4\pi\epsilon_0 r}e^{-r/\lambda_D}.
\label{eq:debye_yukawa_potential}
\end{equation}
For a mobile electron--proton plasma with \(n_e=n_p=n_0\) and
\(T_e=T_p=T\), Eq.~\eqref{eq:debye_multicomponent} gives
\begin{equation}
\lambda_D^2=\frac{\epsilon_0 k_BT}{2n_0e^2}.
\label{eq:debye_two_species}
\end{equation}
If instead only the electron response is retained, with the ions treated as a
fixed neutralizing background, one obtains the common electron Debye length
\begin{equation}
\lambda_{De}^2=\frac{\epsilon_0 k_BT}{n_e e^2}.
\label{eq:debye_electron}
\end{equation}
Thus the factor of two is not universal; it records whether both mobile species
or only the electrons are included. The compact rewriting
\(\lambda_D^2=\epsilon_0 k_BT/(n_{\rm tot}e^2)\) is valid only in the
quasi-neutral symmetric convention \(n_{\rm tot}=n_e+n_p=2n_0\). It is not a
general formula for an uncompensated non-neutral plasma.

Debye-type screening constraints have also been exploited in other cosmological
plasma settings, for example in analyses of magnetic monopole plasmas, where the
existence of large-scale magnetic fields implies strong limits on the allowed screening
length \cite{MedvedevLoeb2017}.

The linear Debye estimate relies on two essential assumptions: (i) the perturbation is small,
$|e\Phi| \ll k_B T$, and (ii) the plasma contains a sufficiently large number of
mobile charges within a Debye sphere to establish local equilibrium. The first of
these assumptions is clearly strained near an astron, where the potential is very
large. Indeed, for the benchmark charge in Eq.~\eqref{eq:fiducial_large_charge}
one has
\begin{equation}
\frac{e\Phi(r)}{k_B T} \gg 1
\end{equation}
over a broad range of distances, so the linear Debye--H\"uckel approximation cannot
be accepted without qualification.

This point can be made explicit by evaluating the dimensionless expansion
parameter of the Debye--H\"uckel approximation,
\begin{equation}
\frac{e\Phi(r)}{k_B T}.
\end{equation}
For an astron of charge \(Q_A\), the electrostatic potential is
\begin{equation}
\Phi(r)=\frac{k_e Q_A}{r},
\end{equation}
so that
\begin{equation}
\frac{e\Phi(r)}{k_B T}
=
\frac{k_e e Q_A}{k_B T\,r}.
\end{equation}
Taking the fiducial large-charge value from Eq.~\eqref{eq:fiducial_large_charge}
and \(T\sim10^6\,\mathrm K\), one finds
\begin{equation}
\frac{e\Phi(r)}{k_B T}
\simeq
\frac{(8.99\times10^9)(1.60\times10^{-19})(4\times10^{32})}
     {(1.38\times10^{-23})(10^6)\,r}
\simeq
\frac{4.2\times10^{40}}{r(\mathrm m)}.
\end{equation}
Thus,
\begin{equation}
\frac{e\Phi(1\,\mathrm{pc})}{k_B T}\simeq 1.4\times10^{24},
\qquad
\frac{e\Phi(1\,\mathrm{Mpc})}{k_B T}\simeq 1.4\times10^{18},
\end{equation}
and even at cosmological distances the ratio remains enormously larger than
unity. The condition
\begin{equation}
\left|\frac{e\Phi}{k_B T}\right|\ll1
\end{equation}
required for the linear Debye--H\"uckel approximation is therefore grossly
violated over the entire range of physically relevant distances. In this sense,
the linear screening formula can at best be used as a rough benchmark,
but not as a self-consistent description of the astron field.

\begin{figure}[t]
\centering
\includegraphics[width=0.54\textwidth]{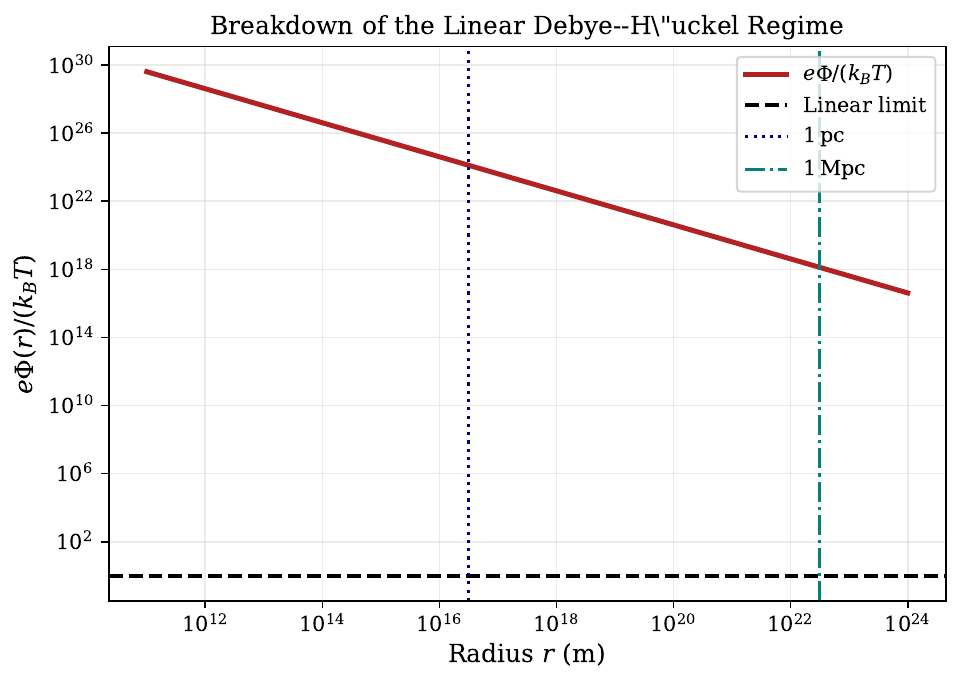}
\caption{Numerical scan of the Debye--H\"uckel linearization parameter
\(e\Phi(r)/(k_B T)\) for the benchmark astron charge in
Eq.~\eqref{eq:fiducial_large_charge}
and \(T=10^6\,\mathrm K\). The horizontal dashed line marks the formal limit of
validity of the linear approximation, while the vertical markers indicate the
scales \(1\,\mathrm{pc}\) and \(1\,\mathrm{Mpc}\). The plot shows that the linear
regime is never reached on physically relevant scales.}
\label{fig:debye_validity}
\end{figure}

Figure~\ref{fig:debye_validity} shows this point in a compact numerical scan. The
ratio \(e\Phi/(k_B T)\) falls only as \(r^{-1}\), so even at parsec and megaparsec
distances it remains vastly above unity for the benchmark charge. The linear Debye
regime is therefore not merely inaccurate near the source; it is absent throughout
the astrophysically relevant domain.

Moreover, the standard estimate need not remain quantitatively reliable when the
plasma departs strongly from local Maxwellian equilibrium. In plasmas with pronounced
suprathermal tails, modeled for example by Kappa distributions, the effective Debye
length can be enhanced relative to its classical value
\begin{pubrev}\cite{PierrardLazar2010,LivadiotisMcComas2013,FahrHeyl2016}\end{pubrev}.
Likewise,
in curved spacetime the effective screening scale acquires gravitational corrections
and redshift effects, as has been shown explicitly for hot non-Abelian plasmas
\cite{AlonsoMonsalveKaiser2023}. Although these analyses do not directly solve the
astron problem, they reinforce the point that screening in an astrophysical or
cosmological environment cannot always be reduced to the flat-space linear formula.

It is nevertheless useful to quote the Debye estimate as a benchmark,
provided the meaning of the estimate is kept clear. The Debye length computed
here is a property of the \emph{ambient intergalactic plasma}, not of the astron
itself. For representative intergalactic conditions,
\begin{equation}
n_e \sim 0.2\, \mathrm{m}^{-3}, \qquad T \sim 10^6 \mathrm{K},
\end{equation}
the one-species electron Debye estimate is
\begin{equation}
\lambda_D \approx 1.5\times 10^{5}\,\mathrm{m},
\end{equation}
namely about \(150\,\mathrm{km}\). This is extremely small compared both with a
megaparsec and with the fiducial astron compactness scale
\(r_g\sim1.5\times10^{15}\,\mathrm m\). Thus, if the linear Debye formula could
be applied naively, it would predict exponential suppression of the electric
field on a scale completely negligible compared with inter-astron separations.

The number of particles in a Debye sphere,
\begin{equation}
N_D = \frac{4\pi}{3} n_e \lambda_D^3,
\end{equation}
has a different meaning. It is a collectivity parameter for the background
plasma: it asks how many ambient particles lie inside a volume of radius
\(\lambda_D\), not how many particles are needed to neutralize the astron.
With the above numbers,
\begin{equation}
N_D \approx 3.1\times 10^{15}.
\end{equation}
Including both electron and proton responses changes these numbers only by factors
of order unity, giving \(\lambda_D\simeq1.1\times10^5\,\mathrm m\) and
\(N_D\simeq1.1\times10^{15}\).
Thus \(N_D\gg1\): the plasma is collective, and the Debye calculation is
not failing because there are only order-one particles in a Debye sphere.

This should be separated from the much larger question of neutralizing the astron.
The number of elementary charges corresponding to the fiducial astron charge is
\begin{equation}
N_Q = \frac{Q_A}{e} \simeq 2.5\times10^{51}
\end{equation}
for the charge in Eq.~\eqref{eq:fiducial_large_charge}.
Clearly \(N_Q\gg N_D\). A single Debye sphere of the unperturbed IGM does not
contain enough opposite charge to compensate the astron. This is not a paradox:
ordinary Debye screening is a linear polarization calculation, not the statement
that one Debye sphere literally contains the full neutralizing charge of an
arbitrarily large source. For the astron, however, the enormous value of \(N_Q\)
is precisely a warning that the linear approximation is not self-consistent near
the source.

\begin{figure}[t]
\centering
\includegraphics[width=0.54\textwidth]{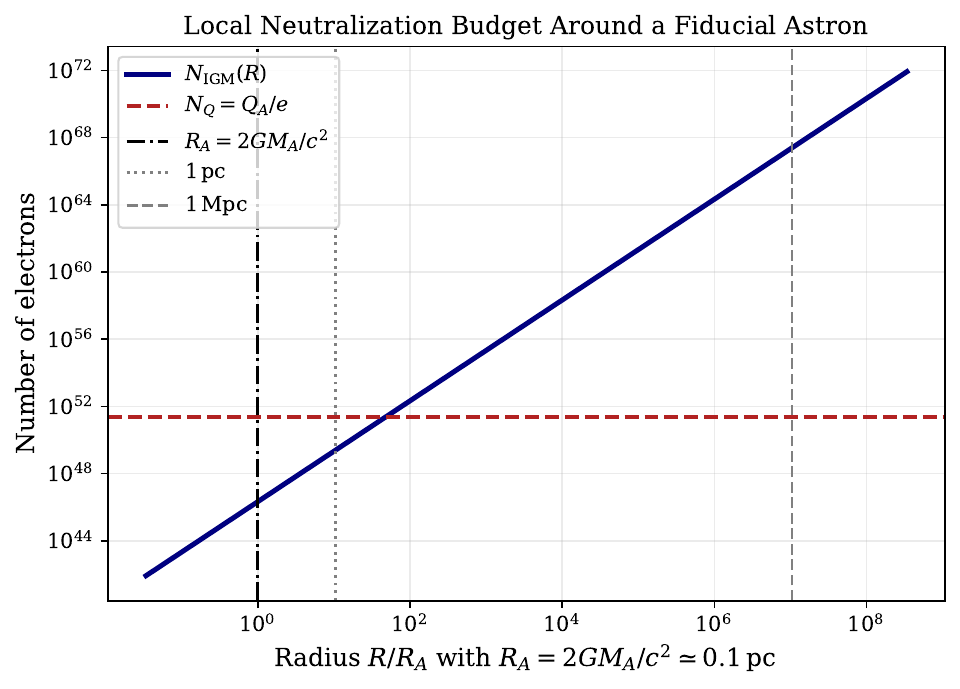}
\caption{Comparison between the number of electrons contained in a sphere of radius
\(R\), \(N_{\rm IGM}(R)=n_e(4\pi/3)R^3\), and the number
\(N_Q=Q_A/e\) required to compensate the benchmark astron charge. The horizontal
axis is normalized to the fiducial astron compactness scale
\(R_A=2GM_A/c^2\simeq0.1\,\mathrm{pc}\), and the plot also marks
\(1\,\mathrm{pc}\) and \(1\,\mathrm{Mpc}\). It shows that complete
neutralization requires a reservoir extending over many astron compactness scales
even before one addresses the kinetic problem of actually transporting those
charges.}
\label{fig:neutralization_counts}
\end{figure}

It is therefore more transparent to ask how large a charge reservoir is needed.
\begin{pubrev}
For the present-day benchmark density in Eq.~\eqref{eq:igm_ne_benchmark}, the
number of ambient electrons enclosed within a sphere of radius \(R\) is given
by Eq.~\eqref{eq:nigm_sphere}.
\end{pubrev}
Equating \(N_{\mathrm{IGM}}(R)\) with \(N_Q\) gives
\begin{equation}
R_Q
\sim
\left(\frac{3N_Q}{4\pi n_e}\right)^{1/3}
\sim 1.4\times10^{17}\,\mathrm m
\sim 5\,\mathrm{pc}.
\end{equation}
Thus a megaparsec-sized region contains vastly more than enough electrons in
principle, but complete compensation of the astron charge would require drawing
on a macroscopic reservoir extending over parsec scales. The real issue is
therefore not a failure of plasma collectivity as measured by \(N_D\), but the
nonlinear transport and rearrangement of a very large amount of charge.
Figure~\ref{fig:neutralization_counts} compares this quantity with the number of
electrons required to compensate the fiducial astron charge. The crossing occurs
only at radii of order several parsecs, so local neutralization is not a compact
near-surface phenomenon even for a realistic mean IGM density.

One may attempt to go beyond the linear regime and consider the full nonlinear Poisson--Boltzmann
equation. In that case the problem becomes intrinsically kinetic and nonlocal. The
timescale for transporting charge over a macroscopic distance $L$ is controlled by
diffusion or drift,
\begin{equation}
t_{\mathrm{screen}} \sim \frac{L^2}{D},
\end{equation}
where $D$ is the diffusion coefficient. Whether this timescale is short or long compared
with cosmological evolution depends on the detailed transport model and cannot be
settled by dimensional analysis alone.

A complementary way to state the issue is that the plasma may be effectively collisionless
on some relevant scales, so the assumptions behind local thermodynamic equilibrium
must be checked rather than assumed. This weakens the direct applicability of linear
screening formulae, but it does not automatically prove the absence of screening.

The linear Debye estimate therefore has a precise role in the argument. It gives
the screening scale predicted by a linear Maxwellian plasma, while the
large-potential regime shows why that scale cannot by itself be promoted to a
self-consistent astron solution. A reliable assessment of the screening problem
requires a nonlinear or kinetic analysis. Appendix~\ref{app:kinetic_screening}
formulates the minimal self-consistent problem in terms of a Vlasov--Poisson or
Vlasov--Fokker--Planck--Poisson description, making explicit which assumptions
must be checked before one can infer either exponential screening or persistence
of the long-range field.

It is useful to summarize the logic in four separate statements. First,
\(N_D\gg1\) means that the intergalactic medium is a collective plasma on the
linear Debye scale; it does not mean that one Debye sphere contains enough
charge to neutralize an astron. Second, the value
\(\lambda_D\sim 150\,\mathrm{km}\) is the linear Debye length of the ambient IGM
plasma and is tiny compared with both the astron compactness scale and a
megaparsec. Third, the charge available in one Debye sphere is
\(eN_D\sim 10^{-3}\,\mathrm C\), enormously smaller than
\(Q_A\sim 10^{32}\,\mathrm C\), so one Debye sphere cannot literally compensate
the astron charge. Fourth, enough opposite charge exists only in a macroscopic
reservoir, of order a few parsecs for the fiducial IGM density. Thus the
screening problem for an astron is not ordinary small-perturbation Debye
screening; it is a nonlinear charge-reservoir and transport problem.

\section{Abundance and Discreteness Constraints}
\label{sec:abundance_constraints}

A population of \(10^{12}M_\odot\) compact objects is not a smooth dark
component on megaparsec scales.  Even before electromagnetic effects are
included, such objects produce gravitational discreteness, Poisson fluctuations,
and dynamical effects that must be confronted with large-scale-structure data.
This section gives only the minimal bookkeeping needed to state the constraint
clearly.

Let \(f_A\) be the fraction of the present matter density carried by astrons.
Using
\begin{equation}
\rho_{m,0}=\Omega_m\rho_{c,0}
\simeq 4\times10^{10}\,M_\odot\,{\rm Mpc}^{-3},
\end{equation}
one finds
\begin{equation}
n_A
=
\frac{f_A\rho_{m,0}}{M_A}
\simeq
4\times10^{-2}\,f_A
\left(\frac{M_A}{10^{12}M_\odot}\right)^{-1}
{\rm Mpc}^{-3}.
\label{eq:astructure_number_density}
\end{equation}
The corresponding mean separation is
\begin{equation}
d_A=n_A^{-1/3}
\simeq
3\,f_A^{-1/3}
\left(\frac{M_A}{10^{12}M_\odot}\right)^{1/3}
{\rm Mpc}.
\label{eq:astructure_separation}
\end{equation}
Thus a separation of order one megaparsec would correspond not to a small
subcomponent, but to a mass density
\begin{equation}
\rho_A\sim 10^{12}\,M_\odot\,{\rm Mpc}^{-3},
\end{equation}
which is roughly twenty-five times the total present matter density.  A
cosmologically allowed population must therefore be substantially more dilute,
or else the fiducial mass must be reduced.

The discreteness also generates a white-noise contribution to the density power
spectrum.  For the astron component itself the shot-noise power is
\begin{equation}
P_A^{\rm shot}=\frac{1}{n_A}.
\end{equation}
If astrons make only a fraction \(f_A\) of the matter, the contribution to the
total matter-density contrast is weighted by \(f_A^2\):
\begin{equation}
P_m^{\rm shot}
\simeq
\frac{f_A^2}{n_A}
=
\frac{f_A M_A}{\rho_{m,0}}
\simeq
25\,f_A
\left(\frac{M_A}{10^{12}M_\odot}\right)
{\rm Mpc}^3.
\label{eq:shot_noise_power}
\end{equation}
The corresponding dimensionless contribution is
\begin{equation}
\Delta_{\rm shot}^2(k)
=
\frac{k^3P_m^{\rm shot}}{2\pi^2}.
\end{equation}
At \(k\sim1\,{\rm Mpc}^{-1}\), Eq.~\eqref{eq:shot_noise_power} gives
\begin{equation}
\Delta_{\rm shot}^2
\sim
1.3\,f_A
\left(\frac{M_A}{10^{12}M_\odot}\right).
\end{equation}
This estimate is only a crude normalization, because the growth of isocurvature
Poisson fluctuations depends on the full cosmological history.  Nevertheless it
shows that the discreteness of ultra-massive compact objects is not a secondary
effect.  Related Poisson-power constraints are standard in discussions of
massive primordial black holes and compact dark-matter subcomponents
\cite{AfshordiMcDonaldSpergel2003,CarrKuhnel2020}.

The abundance constraint is independent of the charge-generation problem.  Even
if a primordial mechanism could create and preserve the fiducial charge, the
population would still have to satisfy bounds from the matter power spectrum,
halo abundances, clustering, lensing, dynamical heating, and the absence of
excessively massive compact objects in observed environments.  A viable astron
cosmology must therefore specify not only \((M_A,Q_A)\), but also the abundance
\(f_A\), spatial distribution, clustering bias, and formation redshift.  Without
that information, claims about cosmic acceleration or early structure formation
remain incomplete.

\section{Astrons and Naked Singularities}

Because the astron scenario invokes extremely large charges for ultra-massive
compact objects, it is natural to ask whether the corresponding geometry is
black-hole-like, merely horizonless, or genuinely nakedly singular. In the
present section we make this distinction explicit in terms of the mass \(M\),
charge \(Q\), and physical radius \(R\) of the object. The key point is that,
once spherical symmetry, vanishing rotation, and an electrovac exterior are
assumed, the Einstein--Maxwell equations fix the exterior geometry from \(M\)
and \(Q\) alone. The remaining ambiguity is not in the exterior metric, but in
how the finite-radius interior is completed.

Under these assumptions the spacetime outside the object is necessarily of
Reissner--Nordstr\"om type,
\begin{equation}
ds^2=f(r)c^2dt^2-f(r)^{-1}dr^2-r^2d\Omega^2,
\qquad
f(r)=1-\frac{2GM}{c^2r}+\frac{Gk_eQ^2}{c^4r^2}.
\label{eq:rn_metric_astron}
\end{equation}
The potential horizons are the zeros of \(f(r)\),
\begin{equation}
r_\pm=\frac{GM}{c^2}\pm
\sqrt{\left(\frac{GM}{c^2}\right)^2-\frac{Gk_eQ^2}{c^4}},
\label{eq:RNhorizons_astron}
\end{equation}
so the outer horizon \(r_+\) is the charged generalization of the Schwarzschild
radius. Indeed, for \(Q\to0\) one recovers \(r_+\to 2GM/c^2\).
For nonzero charge, however, the positive \(Q^2/r^2\) term in \(f(r)\) reduces
the outer horizon radius relative to the Schwarzschild value. Equivalently, the
charge appears as a subtraction inside the square root in
Eq.~\eqref{eq:RNhorizons_astron}: as \(Q\) increases, the square root decreases,
so \(r_+\) moves inward from \(2GM/c^2\) toward \(GM/c^2\) at extremality.

It is useful to introduce the dimensionless parameters
\begin{equation}
\Xi \equiv \frac{k_eQ^2}{GM^2},
\qquad
\eta \equiv \frac{c^2R}{GM}.
\label{eq:xi_eta_def}
\end{equation}
The parameter \(\Xi\) is fixed entirely by the geometry, since
Eq.~\eqref{eq:RNhorizons_astron} may be rewritten as
\begin{equation}
r_\pm=\frac{GM}{c^2}\left(1\pm\sqrt{1-\Xi}\right).
\label{eq:rn_radii_xi}
\end{equation}
Equivalently, the horizon equation \(f(r)=0\) is a quadratic with discriminant
\begin{equation}
\Delta=
\left(\frac{2GM}{c^2}\right)^2-4\frac{Gk_eQ^2}{c^4}
=4\left(\frac{GM}{c^2}\right)^2(1-\Xi).
\end{equation}
The three cases are therefore
\begin{align}
\Xi<1 &: \quad \text{sub-extremal RN with two real horizons}, \\
\Xi=1 &: \quad \text{extremal RN with }r_+=r_-=\frac{GM}{c^2}, \\
\Xi>1 &: \quad \text{super-extremal RN with no horizon.}
\end{align}
This already shows that the existence or absence of a
Reissner--Nordstr\"om horizon is determined entirely by the pair \((M,Q)\), or
equivalently by the single combination \(\Xi\). The radius \(R\) does not enter
the horizon equation at all.
The role of \(R\) becomes relevant only after horizons exist. If \(\Xi\le1\),
the physical question is whether the material body lies inside or outside the
outer horizon \(r_+\). In terms of \(\eta\), the condition \(R<r_+\) becomes
\begin{equation}
\eta < 1+\sqrt{1-\Xi},
\end{equation}
whereas \(R>r_+\) becomes
\begin{equation}
\eta > 1+\sqrt{1-\Xi}.
\end{equation}
Accordingly, for \(\Xi<1\), the two radial regimes are
\begin{align}
\eta<1+\sqrt{1-\Xi}
&: \quad \text{black-hole-like object}, \\
\eta>1+\sqrt{1-\Xi}
&: \quad \text{charged compact object with RN exterior but no horizon covering the surface.}
\end{align}
At extremality, \(\Xi=1\), the critical surface occurs at \(\eta=1\).

By contrast, if \(\Xi>1\), the horizon radii do not exist at all. In that
regime no choice of \(R\) can restore a horizon, because there is no horizon
radius with which \(R\) can be compared. The radius only fixes the size of the
interior region to which the super-extremal RN exterior is matched. In this
precise sense, the pair \((M,Q)\) fixes \(\Xi\), and \(\Xi\) fixes the RN
horizon structure. The radius \(R\) instead fixes the matching of the exterior
geometry to the interior, and it can be compared with \(r_+\) only when
\(\Xi\le1\).

This still does not settle whether the object is a {naked singularity},
because horizonlessness and singularity are not the same notion. For the exact
Reissner--Nordstr\"om solution continued all the way to \(r=0\), the center is
automatically singular: in geometric units the Kretschmann scalar is
\begin{equation}
K=
R_{\mu\nu\rho\sigma}R^{\mu\nu\rho\sigma}
=
\frac{48M^2}{r^6}
-\frac{96MQ^2}{r^7}
+\frac{56Q^4}{r^8},
\end{equation}
so \(K\to\infty\) as \(r\to0\). Therefore an exact super-extremal
Reissner--Nordstr\"om spacetime is automatically a naked singularity. However,
if the same exterior geometry is interpreted only for \(r>R\) and is matched to
a regular finite-radius interior, one obtains instead a horizonless overcharged
compact object. The distinction between these two possibilities is not in the
exterior metric, but in the interior completion.

\subsection{Reissner--Nordstr\"om Exteriors with Regular or Singular Interiors}

The previous discussion shows that the exterior Reissner--Nordstr\"om geometry
by itself does not determine whether the center is regular or singular. To make
this point explicit, we use the standard curvature-coordinate form for a static,
spherically symmetric charged fluid, following the charged-fluid treatment of
Bekenstein and later compact-star applications
\cite{Bekenstein1971ChargedFluid,RayEtAl2003ChargedStars}. In these coordinates
the radial coordinate is the areal radius, and the Einstein--Maxwell constraint
fixes the radial metric coefficient in terms of the enclosed mass \(m(r)\) and
enclosed charge \(q(r)\):
\begin{equation}
ds^2=e^{2\Psi(r)}c^2dt^2-
\left(1-\frac{2Gm(r)}{c^2r}+\frac{Gk_e q(r)^2}{c^4r^2}\right)^{-1}dr^2-r^2d\Omega^2,
\qquad 0\le r<R,
\label{eq:general_charged_interior}
\end{equation}
Here \(m(r)\) and \(q(r)\) are the enclosed mass and enclosed charge functions,
and \(\Psi(r)\) is the gravitational redshift function. At this stage
\(\Psi(r)\) is left general because no microscopic equation of state or charge
profile has yet been specified. It is not, however, an arbitrary physical
profile in a complete interior solution. Once the matter density, pressure, and
charge density are chosen, the Einstein--Maxwell equations determine
\(d\Psi/dr\) through the charged TOV equation, Eq.~\eqref{eq:charged_TOV_phieq}.
The remaining additive constant in \(\Psi\) is only the normalization of the
time coordinate, and is fixed by matching \(g_{tt}\) to the exterior metric at
\(r=R\). This is the usual charged generalization of the interior
Schwarzschild/TOV parametrization; for \(q(r)=0\) it reduces to
\(g_{rr}^{-1}=1-2Gm(r)/(c^2r)\).
For \(r>R\) the exterior remains
\begin{equation}
ds^2=f_{\rm RN}(r)c^2dt^2-f_{\rm RN}(r)^{-1}dr^2-r^2d\Omega^2,
\qquad
f_{\rm RN}(r)=1-\frac{2GM}{c^2r}+\frac{Gk_eQ^2}{c^4r^2}.
\end{equation}
Continuity of the metric at the matching surface \(r=R\) requires
\begin{equation}
m(R)=M,\qquad q(R)=Q,\qquad e^{2\Psi(R)}=f_{\rm RN}(R).
\end{equation}
If one also imposes continuity of the extrinsic curvature, the matching is
smooth; otherwise the jump is interpreted as a thin shell with surface
stress-energy through the Israel junction conditions
\begin{pubrev}\cite{Darmois1927,Israel1966Junction,MarsSenovilla1993Junction,Poisson2004Toolkit}\end{pubrev}.

Regularity at the center follows if the enclosed mass and charge behave as
\begin{equation}
m(r)\sim m_3 r^3,\qquad q(r)\sim q_3 r^3,\qquad \Psi(r)\to\Psi_0
\qquad (r\to0).
\label{eq:regular_center_conditions}
\end{equation}
Indeed, under these conditions one has
\begin{equation}
\frac{m(r)}{r}\sim r^2,
\qquad
\frac{q(r)^2}{r^2}\sim r^4,
\end{equation}
so the metric coefficients remain finite at the origin. The electric field is
then also regular, since
\begin{equation}
E(r)\sim \frac{k_e q(r)}{r^2}\sim r.
\end{equation}
Thus an object with \(R>r_+\) can perfectly well be a finite charged compact
body with a regular center, in direct analogy with an ordinary star.

By contrast, singular behavior arises if the interior is chosen so that
\begin{equation}
m(r)\to m_0\neq0
\qquad\text{or}\qquad
q(r)\to q_0\neq0
\qquad (r\to0),
\end{equation}
because then the metric develops the schematic behavior
\begin{equation}
1-\frac{2Gm(r)}{c^2r}+\frac{Gk_e q(r)^2}{c^4r^2}
\sim
1-\frac{2Gm_0}{c^2r}+\frac{Gk_e q_0^2}{c^4r^2},
\end{equation}
which is singular at \(r=0\). The exact Reissner--Nordstr\"om continuation to
the center is precisely of this type.

Two simple examples illustrate the distinction. A regular thin-shell model is
obtained by taking the region \(r<R\) to be flat spacetime, written with a
constant lapse chosen to match the exterior time normalization at the shell:
\begin{equation}
ds^2_{\rm in}=f_{\rm RN}(R)c^2dt^2-dr^2-r^2d\Omega^2,
\qquad r<R,
\end{equation}
where \(f_{\rm RN}(R)>0\) is assumed so that the surface is timelike. This is
flat because the constant rescaling \(T=\sqrt{f_{\rm RN}(R)}\,t\) brings the
interior metric to the Minkowski form
\begin{equation}
ds^2_{\rm in}=c^2dT^2-dr^2-r^2d\Omega^2 .
\end{equation}
Thus ``flat interior'' means that there is no volume mass density, no volume
charge density, and no electromagnetic field for \(r<R\). The mass and charge
are instead supported at the boundary: the charge appears as a surface charge
on the shell, while the jump in extrinsic curvature is balanced by the shell
stress-energy. In this case the center is manifestly regular, although the
matching surface is a material thin shell rather than a smooth fluid boundary.
A smooth regular interior may instead be
modeled by the ansatz
\begin{equation}
m(r)=M\left(\frac{r}{R}\right)^3,
\qquad
q(r)=Q\left(\frac{r}{R}\right)^3,
\label{eq:smooth_regular_ansatz}
\end{equation}
with \(\Psi(r)\) chosen regular and satisfying the matching condition at
\(r=R\). In a complete fluid model \(\Psi(r)\) would be determined by the
charged TOV equation once the equation of state and charge density are fixed;
in this illustrative ansatz it is only required to remain finite at the center
and to match the exterior lapse at the surface. Then
\begin{equation}
g_{rr}^{-1}
=
1-\frac{2GM}{c^2R^3}r^2+\frac{Gk_eQ^2}{c^4R^6}r^4,
\end{equation}
which is manifestly finite at the origin. Thus the same
Reissner--Nordstr\"om exterior for \(r>R\) may correspond either to a regular
finite-radius charged object or to a naked singularity. The distinction is not
fixed by the exterior geometry alone, but by the choice of interior
completion.

\subsection{Summary: Astrons with and without Horizons}

It is useful to make the previous discussion even more explicit by writing the
full spacetime as a piecewise metric. In the present static, spherically
symmetric setting, the natural form is
\begin{equation}
ds^2=
\begin{cases}
ds^2_{\rm int}, & 0\le r<R,\\[1mm]
ds^2_{\rm RN}, & r>R,
\end{cases}
\qquad
ds^2_{\rm RN}=f_{\rm RN}(r)c^2dt^2-f_{\rm RN}(r)^{-1}dr^2-r^2d\Omega^2.
\label{eq:piecewise_metric_astron}
\end{equation}
The qualitative nature of the configuration is then fixed by \(\Xi\), defined in
Eq.~\eqref{eq:xi_eta_def}, and by the position of \(R\) relative to the outer RN
horizon \(r_+\), whenever the latter exists.

If \(\Xi<1\), the vacuum Reissner--Nordstr\"om geometry possesses the outer
horizon given by the positive branch of
\begin{pubrev}Eq.~\eqref{eq:rn_radii_xi}.\end{pubrev}
This formula makes explicit why the RN outer horizon is smaller than the
Schwarzschild radius: since \(0<\sqrt{1-\Xi}<1\) for \(0<\Xi<1\), one has
\(GM/c^2<r_+<2GM/c^2\). The limit \(\Xi=0\) gives the Schwarzschild value,
while \(\Xi=1\) gives the extremal radius \(r_+=GM/c^2\).
If in addition
\begin{equation}
R<r_+,
\end{equation}
then the matching surface lies inside the outer horizon and the full spacetime
is black-hole-like. Two subcases are then possible. The singular realization is
\begin{pubrev}
the exact RN black hole obtained by extending Eq.~\eqref{eq:rn_metric_astron}
to the origin; its center is singular. The regular realization instead uses
the piecewise construction in Eq.~\eqref{eq:piecewise_metric_astron}, with the
interior metric of Eq.~\eqref{eq:general_charged_interior} and \(R<r_+\).
\end{pubrev}
In this case the horizon is present, but the central region is
regular rather than singular.

If \(\Xi<1\) but
\begin{equation}
R>r_+,
\end{equation}
the situation is different. The surface of the object lies outside the radius
at which the vacuum RN metric would have developed an outer horizon. Since the
region \(0\le r<R\) is not vacuum but is instead described by the interior
metric \(ds^2_{\rm int}\), the full spacetime need not contain any horizon at
all. This is exactly analogous to the Sun: although the exterior Schwarzschild
solution has a would-be horizon at \(2GM/c^2\), the physical star ends long
before that radius is reached, and the interior fluid solution replaces the
vacuum metric. In the astron case the corresponding horizonless regular
configuration is again given by Eq.~\eqref{eq:piecewise_metric_astron}, but now
with \(R>r_+\).

Finally, if \(\Xi>1\), the exterior Reissner--Nordstr\"om geometry is
super-extremal and no RN horizon exists anywhere. Here again two conceptually
different possibilities arise. If one takes the exact RN geometry all the way
\begin{pubrev}
to the origin, Eq.~\eqref{eq:rn_metric_astron} with \(\Xi>1\) describes a
naked singularity. If instead one matches the same exterior to the regular
finite-radius interior in Eq.~\eqref{eq:piecewise_metric_astron}, one has a
horizonless overcharged compact object rather than a naked
\end{pubrev}
singularity.

The four cases may therefore be summarized as follows:
\begin{align}
\Xi<1,\ R<r_+
&: \quad \text{horizon present}, \\
\Xi<1,\ R>r_+
&: \quad \text{finite charged object, no horizon in the full spacetime}, \\
\Xi>1,\ \text{exact RN}
&: \quad \text{naked singularity}, \\
\Xi>1,\ \text{regular interior}
&: \quad \text{horizonless overcharged compact object}.
\end{align}
In this way the role of the radius \(R\) becomes completely transparent: it
does not alter the exterior RN geometry, but it does determine whether the full
spacetime includes only the vacuum RN region or instead replaces the central
part of that geometry by a regular interior.

\subsection{Charged TOV Equations for Regular Astrons}

If one wishes to model an astron as a regular static charged object rather than
as an exact vacuum Reissner--Nordstr\"om singularity, the appropriate interior
equations are the Einstein--Maxwell generalization of the
Tolman--Oppenheimer--Volkoff system. This description is meaningful for the
regular configurations discussed above, namely the horizonless case \(R>r_+\)
and the super-extremal case \(\Xi>1\) with a regular interior. By contrast, it
does not describe the exact RN naked singularity, since that solution is
electrovac all the way to \(r=0\), nor is it the natural description of a
fully collapsed black hole interior.

The metric form used for the interior is not an additional dynamical assumption.
It is the standard parametrization of a static, spherically symmetric
Einstein--Maxwell configuration in terms of the enclosed gravitational mass and
the enclosed electric charge. This is the usual starting point for the
Einstein--Maxwell generalization of hydrostatic equilibrium for charged fluid
spheres \cite{Bekenstein1971ChargedFluid,RayEtAl2003ChargedStars}. The most
general static, spherically symmetric line
element can first be written as
\begin{equation}
ds^2=e^{2\Psi(r)}c^2dt^2-e^{2\Lambda(r)}dr^2-r^2d\Omega^2,
\label{eq:charged_TOV_metric}
\end{equation}
where the choice of \(r\) as the area radius fixes the angular part to
\(-r^2d\Omega^2\). The function \(\Psi(r)\) is the gravitational redshift
function: it controls clock rates through \(g_{tt}=e^{2\Psi}c^2\). The
function \(\Lambda(r)\) determines the radial geometry.

In the charged problem it is useful to introduce the charge \(q(r)\) enclosed
inside the sphere of area \(4\pi r^2\). This definition is fixed by Maxwell's
equation and gives Gauss' law in curved spacetime,
\begin{equation}
q(r)=4\pi\int_0^r \rho_e(\bar r)\,e^{\Lambda(\bar r)}\bar r^2\,d\bar r .
\label{eq:q_enclosed_integral}
\end{equation}
The \(tt\) Einstein equation may then be integrated once. The result is that the
radial metric coefficient can be written as
\begin{equation}
e^{-2\Lambda(r)}
=
1-\frac{2Gm(r)}{c^2r}+\frac{Gk_e q(r)^2}{c^4r^2},
\label{eq:charged_mass_function}
\end{equation}
where \(m(r)\) is the Misner--Sharp mass generalized to include the
electromagnetic contribution. This definition is chosen so that, outside the
surface where \(m(r)=M\) and \(q(r)=Q\), Eq.~\eqref{eq:charged_mass_function}
reduces exactly to the exterior Reissner--Nordstr\"om metric. It also has the
correct neutral limit: when \(q(r)=0\), it becomes the usual interior
Schwarzschild/TOV parametrization \(e^{-2\Lambda}=1-2Gm/(c^2r)\).

\begin{pubrev}
With these definitions the interior line element is precisely the metric in
Eq.~\eqref{eq:general_charged_interior}.
\end{pubrev}
This is a convenient rewriting of the general static spherical metric,
not a separate choice of equation of state or matter model. Once this form is
adopted, the Einstein--Maxwell equations reduce to a closed system for
\(m(r)\), \(q(r)\), \(\Psi(r)\), and the fluid variables \(\rho(r)\) and
\(P(r)\).

Let the matter sector be modeled as a perfect fluid with proper energy density
\(\rho\), pressure \(P\), and proper charge density \(\rho_e\). Then the charge
function satisfies Gauss' law in curved spacetime,
\begin{equation}
\frac{dq}{dr}=4\pi r^2 \rho_e\,e^{\Lambda(r)}.
\label{eq:charged_TOV_qeq}
\end{equation}
The Einstein equations give the mass equation
\begin{equation}
\frac{dm}{dr}
=
4\pi r^2 \rho
+\frac{k_e q}{c^2r}\frac{dq}{dr},
\label{eq:charged_TOV_meq}
\end{equation}
where the second term represents the contribution of the electrostatic field to
the total mass function.

The remaining independent Einstein equation yields
\begin{equation}
\frac{d\Psi}{dr}
=
\frac{
\dfrac{Gm(r)}{c^2r^2}
+\dfrac{4\pi G}{c^4}\,rP
-\dfrac{Gk_e q(r)^2}{c^4r^3}
}
{
1-\dfrac{2Gm(r)}{c^2r}
+\dfrac{Gk_e q(r)^2}{c^4r^2}
}.
\label{eq:charged_TOV_phieq}
\end{equation}
Hydrostatic equilibrium is obtained from local conservation of total
stress-energy, equivalently from the Euler equation with the Lorentz-force
term:
\begin{equation}
\frac{dP}{dr}
=
-(\rho c^2+P)\frac{d\Psi}{dr}
+\rho_e E\,e^{\Lambda(r)}.
\label{eq:charged_TOV_euler}
\end{equation}
Using Eq.~\eqref{eq:charged_TOV_qeq} together with
\(E(r)=k_e q(r)/r^2\), this may be rewritten as
\begin{equation}
\frac{dP}{dr}
=
-(\rho c^2+P)\frac{d\Psi}{dr}
+\frac{k_e q(r)}{4\pi r^4}\frac{dq}{dr}.
\label{eq:charged_TOV_pressure1}
\end{equation}
Substituting Eq.~\eqref{eq:charged_TOV_phieq} then gives the charged TOV
equation in explicit form,
\begin{equation}
\frac{dP}{dr}
=
-(\rho c^2+P)
\frac{
\dfrac{Gm(r)}{c^2r^2}
+\dfrac{4\pi G}{c^4}\,rP
-\dfrac{Gk_e q(r)^2}{c^4r^3}
}
{
1-\dfrac{2Gm(r)}{c^2r}
+\dfrac{Gk_e q(r)^2}{c^4r^2}
}
+\frac{k_e q(r)}{4\pi r^4}\frac{dq}{dr}.
\label{eq:charged_TOV_final}
\end{equation}

Equations~\eqref{eq:charged_TOV_qeq}, \eqref{eq:charged_TOV_meq}, and
\eqref{eq:charged_TOV_final}, supplemented by an equation of state
\(P=P(\rho)\) and by regularity conditions at \(r=0\), define the charged
TOV problem for a regular astron interior. The boundary conditions are
\begin{equation}
m(0)=0,\qquad q(0)=0,\qquad P(0)=P_c,
\end{equation}
and the surface \(r=R\) is determined by
\begin{equation}
P(R)=0.
\end{equation}
The resulting solution is then matched to the exterior RN geometry with
\begin{equation}
m(R)=M,\qquad q(R)=Q.
\end{equation}

Thus, if astrons are interpreted as regular finite-radius charged compact
objects, the Einstein--Maxwell TOV system provides the natural interior model.
If instead one insists on the exact RN geometry all the way to the center, then
no interior TOV description exists: one is dealing directly with a singular
electrovac spacetime.

\begin{figure}[t]
\centering
\includegraphics[width=0.86\textwidth]{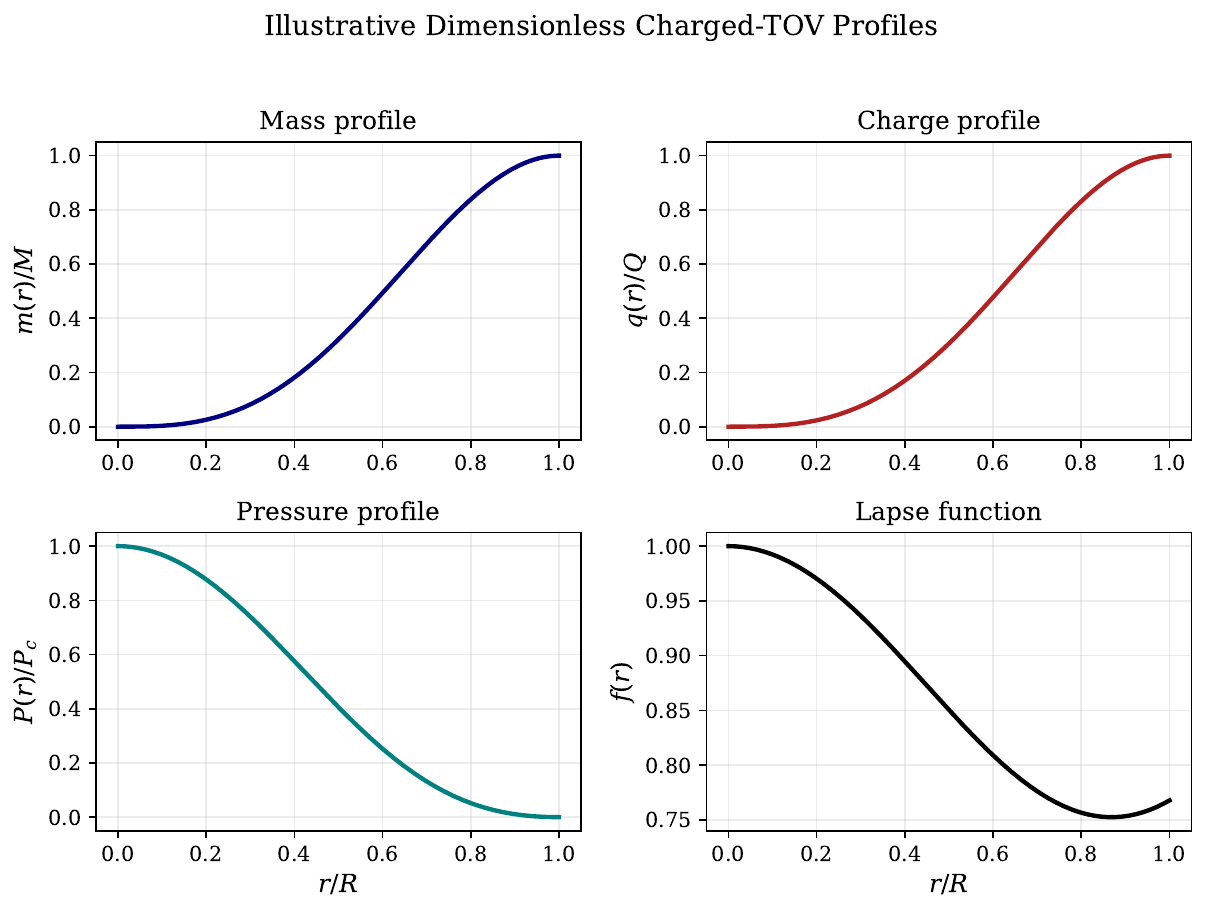}
\caption{Single four-panel summary of an illustrative dimensionless charged-TOV
integration for a regular
Einstein--Maxwell interior with polytropic equation of state \(P=K\rho^\Gamma\),
\(\Gamma=2\), \(K=0.08\), and charge density \(\rho_e=\alpha\rho\) with
\(\alpha=0.08\), in units \(G=c=k_e=1\). {\color{black}The central density is
\(\rho_c=1\), so \(P_c=K\rho_c^\Gamma=0.08\).  The integration reaches
\(P(R)=0\) at \(R\simeq0.304\), with \(M\simeq0.0353\) and
\(Q\simeq0.00311\), giving \(\Xi\simeq7.8\times10^{-3}\).  The lower-right
panel shows the temporal lapse \(e^{\Psi(r)}\), with the additive constant in
\(\Psi\) fixed by the surface matching condition
\(e^{2\Psi(R)}=1-2M/R+Q^2/R^2\simeq0.768\).} The figure is meant as
a qualitative demonstration that a regular charged interior can be matched to an
RN exterior without introducing a central singularity.}
\label{fig:tov_profiles}
\end{figure}

To illustrate what such a regular interior looks like, Fig.~\ref{fig:tov_profiles}
shows a dimensionless charged-TOV integration for a simple polytropic equation of
state \(P=K\rho^\Gamma\) with \(\Gamma=2\), \(K=0.08\), and a proportional charge
density \(\rho_e=\alpha\rho\) with \(\alpha=0.08\), all in units \(G=c=k_e=1\).
The purpose of the plot is not to present a calibrated astron model, but to make
the structure of a regular Einstein--Maxwell interior explicit: the mass and charge
functions rise monotonically from the center, the pressure falls smoothly to zero
at the surface, and the lapse remains positive throughout the interior.

This discussion is especially transparent for the benchmark values proposed in
Ref.~\cite{Frampton2022EAU}: the fiducial mass in
Eq.~\eqref{eq:fiducial_astron_mass} together with the charge in
Eq.~\eqref{eq:fiducial_large_charge}. For this branch
one finds
\begin{equation}
\Xi_A=\frac{k_eQ_A^2}{GM_A^2}\simeq 5.4.
\label{xi}
\end{equation}
This is well above the extremal threshold \(\Xi=1\), so for the Frampton
benchmark the horizon question is already settled by \(M_A\) and \(Q_A\): no
Reissner--Nordstr\"om horizon exists, independently of \(R\). In this regime
the radius no longer decides whether the astron is a black hole; it decides
only how one interprets the interior. By contrast, the ordinary
accretion-saturation charges derived in Secs.~2--4 correspond to \(\Xi\ll1\),
so they lie extremely far from the super-extremal threshold.

\begin{figure}[t]
\centering
\includegraphics[width=0.92\textwidth]{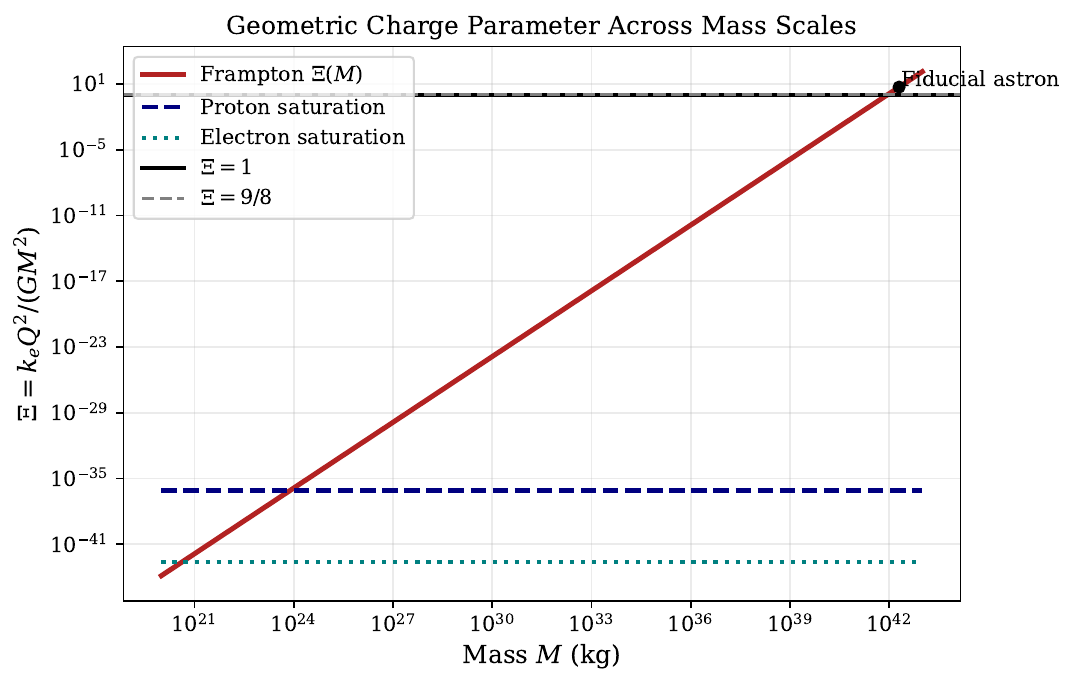}
\caption{The dimensionless geometric charge parameter
\(\Xi=k_eQ^2/(GM^2)\), defined in Eq.~\eqref{eq:xi_eta_def}, as a function of
mass after imposing a charge prescription \(Q=Q(M)\). Thus \(Q\) is included
implicitly in each curve. The Frampton extrapolation uses
\(Q_F=10^{-52}M^2\), giving \(\Xi_F\propto M^2\), while the ordinary proton- and
electron-limited saturation charges scale as \(Q_{\rm sat}\propto M\), giving
nearly constant \(\Xi\). The left panel shows the full dynamic range, while the
right panel zooms into the threshold region where the extremality line
\(\Xi=1\), the photon-sphere threshold \(\Xi=9/8\), and the Frampton fiducial
point would otherwise be compressed together on the logarithmic scale. The figure
shows directly that ordinary saturation remains sub-extremal by an enormous
margin, whereas the Frampton branch enters the super-extremal regime in the mass
range of interest for astrons.}
\label{fig:xi_vs_mass}
\end{figure}

This distinction is displayed quantitatively in Fig.~\ref{fig:xi_vs_mass},
which shows the geometric parameter \(\Xi\), defined in
Eq.~\eqref{eq:xi_eta_def}, for the different
charge prescriptions as a function of mass. The charge is therefore not absent
from the plot; rather, each curve is obtained by first choosing a mass--charge
relation \(Q(M)\) and then substituting it into
\(\Xi=k_eQ^2/(GM^2)\). For the ordinary saturation estimates,
\(Q_{\rm sat}^{(i)}\sim GMm_i/(k_e e)\), the charge scales linearly with \(M\),
so the factor of \(M^2\) in \(Q^2\) cancels the \(M^2\) in the denominator of
\(\Xi\). The corresponding \(\Xi\) values are therefore essentially constant
and extremely small. The Frampton branch, by contrast, uses
\(Q_F(M)=10^{-52}M^2\). Substitution gives
\begin{equation}
\Xi_F(M)
=
\frac{k_e}{G}\,10^{-104}M^2
=
1.35\times10^{-84}\left(\frac{M}{\mathrm{kg}}\right)^2 .
\end{equation}
Equivalently, in solar-mass units,
\begin{equation}
\Xi_F(M)
=
5.32\times10^{-24}
\left(\frac{M}{M_\odot}\right)^2 .
\end{equation}
Thus the fiducial value \(M_A=10^{12}M_\odot\) gives
\begin{equation}
\Xi_F(M_A)\simeq 5.32 .
\end{equation}
The two relevant thresholds are reached at
\begin{equation}
\Xi_F=1:
\qquad
M\simeq 8.62\times10^{41}\,\mathrm{kg}
\simeq 4.33\times10^{11}M_\odot ,
\end{equation}
and
\begin{equation}
\Xi_F=\frac{9}{8}:
\qquad
M\simeq 9.14\times10^{41}\,\mathrm{kg}
\simeq 4.60\times10^{11}M_\odot .
\end{equation}
For comparison, the ordinary saturation branches give the mass-independent
values. In SI units the corresponding charge laws are
\begin{equation}
Q_{\rm sat}^{(p)}
\simeq
7.75\times10^{-29}
\left(\frac{M}{\mathrm{kg}}\right)\mathrm{C},
\qquad
Q_{\rm sat}^{(e)}
\simeq
4.22\times10^{-32}
\left(\frac{M}{\mathrm{kg}}\right)\mathrm{C}.
\end{equation}
Substitution into \(\Xi\) gives
\begin{equation}
\Xi_{\rm sat}^{(p)}
=
\frac{Gm_p^2}{k_e e^2}
\simeq 8.09\times10^{-37},
\qquad
\Xi_{\rm sat}^{(e)}
=
\frac{Gm_e^2}{k_e e^2}
\simeq 2.40\times10^{-43}.
\end{equation}
These numbers explain the visual structure of the plot: the ordinary saturation
curves are horizontal and many orders of magnitude below extremality, whereas
the Frampton curve rises as \(M^2\), appearing as a straight line of slope two on
the log--log plot. It crosses both the extremality bound \(\Xi=1\) and the
photon-sphere threshold \(\Xi=9/8\) in the range relevant to the fiducial astron
benchmark. In this sense the figure makes explicit that the ``large-charge''
astron proposal is not a small deformation of an ordinary charged compact
object, but a transition into a different geometric sector of the
Einstein--Maxwell system.

\begin{figure}[t]
\centering
\includegraphics[width=0.64\textwidth]{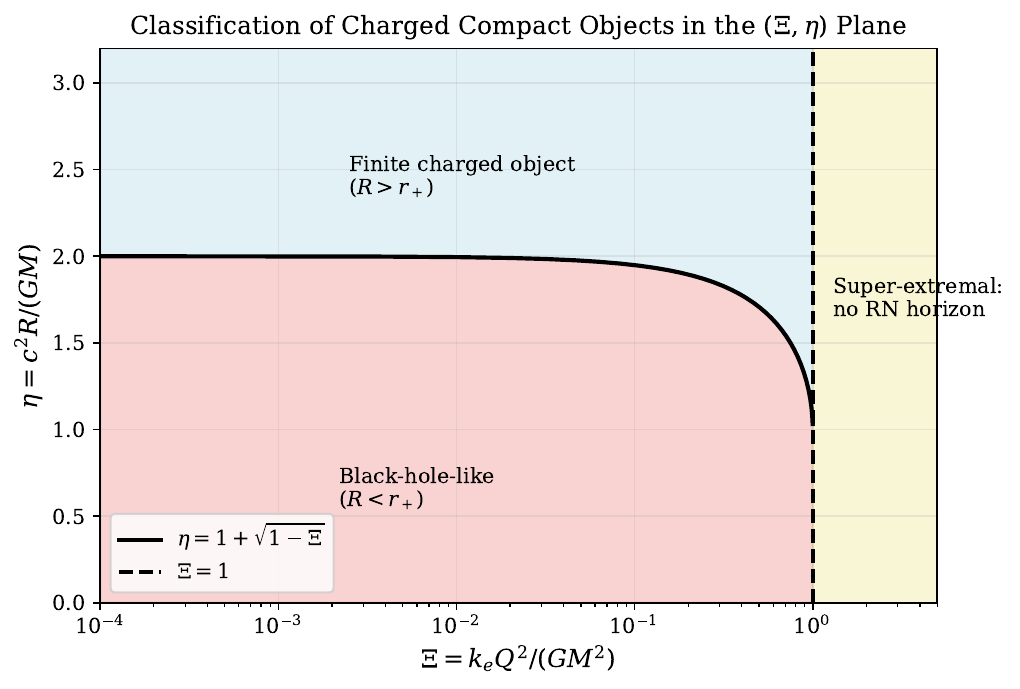}
\caption{Classification of charged compact objects in the
\((\Xi,\eta)\) plane, with both parameters defined in
Eq.~\eqref{eq:xi_eta_def}. For
\(\Xi<1\), the solid curve \(\eta=1+\sqrt{1-\Xi}\) separates black-hole-like
configurations \((R<r_+)\) from finite charged objects with surfaces outside the
outer RN horizon \((R>r_+)\). For \(\Xi>1\), the spacetime is super-extremal
and no RN horizon exists, independently of the value of \(R\).}
\label{fig:xi_eta}
\end{figure}

It is also useful to display separately the roles of the dimensionless parameters
\(\Xi\) and \(\eta\), defined in Eq.~\eqref{eq:xi_eta_def}. Figure~\ref{fig:xi_eta}
shows the resulting classification in the \((\Xi,\eta)\) plane. For
\(\Xi<1\), the boundary \(\eta=1+\sqrt{1-\Xi}\) separates objects whose
surfaces lie inside the outer RN horizon from those whose surfaces lie outside
it. Once \(\Xi>1\), however, no Reissner--Nordstr\"om horizon exists at all,
and varying \(R\) cannot restore one. This is precisely the sense in which
\((M,Q)\) control the horizon structure, while \(R\) controls only the interior
matching whenever a horizon exists.

This point also clarifies how astrons relate to other classes of horizonless
compact objects. A first relevant comparison is with gravastar-type models in
the sense of Mazur and Mottola \cite{MazurMottola2004}. A gravastar replaces
the central singularity by a regular interior, typically of de Sitter type,
matched through a shell to an exterior Schwarzschild geometry; charged
generalizations can likewise be matched to a Reissner--Nordstr\"om exterior.
Gravastars therefore show explicitly that a horizonless object need not be
singular. For astrons, this is conceptually useful because the same
super-extremal RN exterior can be interpreted either as a naked singularity or
as the exterior field of a regular finite-radius object.

If rotation is included, the relevant comparison shifts from
Reissner--Nordstr\"om to Kerr and Kerr--Newman. In those spacetimes horizons
disappear if the spin or the combined spin--charge parameters exceed their
extremal limits, so over-rotating Kerr and over-extremal Kerr--Newman
solutions are also naked singularities \cite{Kerr1963,NewmanEtAl1965}. These
are close in spirit to the astron picture because Kerr--Newman shares the same
Einstein--Maxwell origin as Reissner--Nordstr\"om while adding angular
momentum. The static RN discussion given here should therefore be read as the
simplest charge-dominated limit of a more general charged and rotating compact
object.

Several standard naked-singularity metrics are often used as comparison
geometries, including the Janis--Newman--Winicour solution, the
Zipoy--Voorhees metric, and the Tomimatsu--Sato family. They are useful
mathematical reference points, but they are not the mechanism at work in the
astron model: JNW is sourced by a scalar field, while Zipoy--Voorhees and
Tomimatsu--Sato are vacuum multipole
geometries rather than electrically charged compact-object exteriors
\begin{pubrev}\cite{JanisNewmanWinicour1968,Wyman1981,Zipoy1966,Voorhees1970,TomimatsuSato1972,TomimatsuSato1973}\end{pubrev}.

The main lesson is therefore simple. Under spherical symmetry, no rotation, and
an electrovac exterior, an astron is necessarily described outside the source by
Reissner--Nordstr\"om geometry, and the control parameter is
\(\Xi\) as defined in Eq.~\eqref{eq:xi_eta_def}. Once \(\Xi\) is fixed, the radius \(R\) tells us whether
a sub-extremal object lies inside or outside its outer horizon; in the
super-extremal case it cannot change the absence of horizons. Accordingly, if
one adopts the large Frampton charge, the astron is naturally driven into the
super-extremal Einstein--Maxwell regime. Whether one then calls it a naked
singularity or a horizonless overcharged compact object depends on whether the
interior is taken to be singular or regular. In that precise sense, the astron
scenario sits at the intersection between the Reissner--Nordstr\"om branch of
naked singularities and the wider class of horizonless compact alternatives,
including gravastar-like completions.

\section{Extension to Rotating Astrons: Kerr--Newman Geometry and Dynamics}

The analysis presented so far has been restricted to static, spherically symmetric configurations, for which the exterior geometry is of Reissner--Nordstr\"om type. A realistic astrophysical or primordial compact object, however, is expected to carry angular momentum. The natural generalization is therefore to stationary, axisymmetric configurations described by the Kerr--Newman solution of the Einstein--Maxwell equations. This extension is not merely technical: rotation modifies the extremality condition, the capture dynamics, and the stability properties of the charge sector in a qualitatively nontrivial way.

The Kerr--Newman metric in Boyer--Lindquist coordinates is
\begin{equation}
ds^2 = \frac{\Delta}{\Sigma}(c\,dt - a \sin^2\theta\, d\phi)^2 
- \frac{\Sigma}{\Delta}dr^2 - \Sigma\, d\theta^2 
- \frac{\sin^2\theta}{\Sigma}\left[(r^2 + a^2)d\phi - a c\,dt\right]^2,
\label{eq:KN_metric}
\end{equation}
where
\begin{equation}
\Sigma = r^2 + a^2 \cos^2\theta, 
\qquad 
\Delta = r^2 - \frac{2GM}{c^2}r + \frac{Gk_e Q^2}{c^4} + a^2,
\label{eq:KN_functions}
\end{equation}
and the rotation parameter is defined as
\begin{equation}
a = \frac{J}{Mc}.
\label{eq:spin_parameter}
\end{equation}

The electromagnetic potential is correspondingly given by
\begin{equation}
A_\mu dx^\mu = - \frac{k_e Q r}{\Sigma} \left( c\,dt - a \sin^2\theta\, d\phi \right).
\label{eq:KN_potential}
\end{equation}

The horizon structure is determined by the zeros of $\Delta$,
\begin{equation}
r_\pm = \frac{GM}{c^2} \pm \sqrt{\left(\frac{GM}{c^2}\right)^2 - \left(a^2 + \frac{Gk_e Q^2}{c^4}\right)}.
\label{eq:KN_horizons}
\end{equation}
The condition for the existence of horizons is therefore
\begin{equation}
\left(\frac{GM}{c^2}\right)^2 \ge a^2 + \frac{Gk_e Q^2}{c^4}.
\label{eq:KN_extremality_raw}
\end{equation}
It is convenient to introduce the dimensionless parameters
\begin{equation}
\Xi_Q \equiv \frac{k_e Q^2}{GM^2}, 
\qquad 
\Xi_J \equiv \frac{c^2 J^2}{G^2 M^4},
\label{eq:Xi_parameters}
\end{equation}
in terms of which the extremality condition becomes
\begin{equation}
\Xi_Q + \Xi_J \le 1.
\label{eq:KN_extremality}
\end{equation}

This result generalizes the static Reissner--Nordstr\"om classification. In particular,
\begin{align}
\Xi_Q + \Xi_J < 1 &: \quad \text{sub-extremal configuration with horizons}, \label{eq:KN_subext} \\
\Xi_Q + \Xi_J = 1 &: \quad \text{extremal configuration}, \label{eq:KN_ext} \\
\Xi_Q + \Xi_J > 1 &: \quad \text{super-extremal configuration without horizons}. \label{eq:KN_superext}
\end{align}

Thus rotation and charge compete in determining the global structure of the spacetime. For fixed mass, the presence of angular momentum reduces the maximal charge compatible with a horizon. In particular, the large-charge branch discussed in the static case is driven even more deeply into the super-extremal regime once rotation is included.

\begin{figure}[t]
\centering
\includegraphics[width=0.54\textwidth]{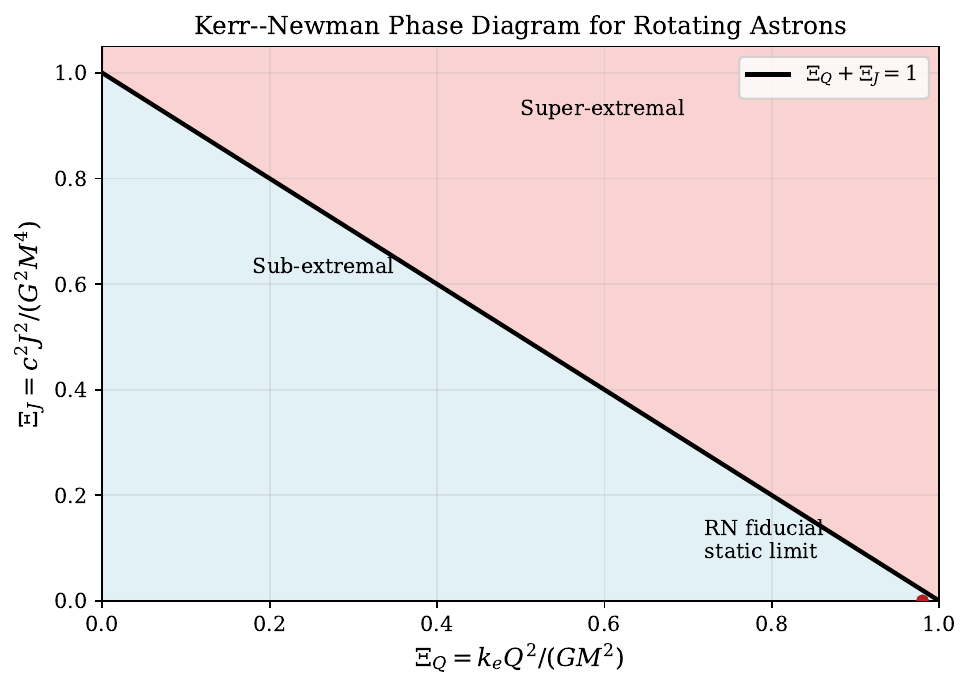}
\caption{Kerr--Newman extremality diagram in terms of the dimensionless charge
and spin parameters \(\Xi_Q=k_eQ^2/(GM^2)\) and \(\Xi_J=c^2J^2/(G^2M^4)\). The
line \(\Xi_Q+\Xi_J=1\) separates sub-extremal configurations with horizons from
super-extremal horizonless configurations. The static RN limit lies on the
\(\Xi_J=0\) axis. {\color{black}The point \((1,0)\) is the extremal RN
boundary, while the fiducial large-charge astron benchmark lies at
\((\Xi_Q,\Xi_J)\simeq(5.4,0)\), deep in the super-extremal region.}
The figure makes explicit that angular momentum further reduces the maximal
charge compatible with a horizon.}
\label{fig:kn_phase}
\end{figure}

This structure is summarized in Fig.~\ref{fig:kn_phase}, which displays the
Kerr--Newman extremality condition in the \((\Xi_Q,\Xi_J)\) plane. The boundary
\(\Xi_Q+\Xi_J=1\) separates the sub-extremal sector from the horizonless one.
The static Reissner--Nordstr\"om analysis corresponds to the axis \(\Xi_J=0\);
once rotation is added, the admissible charge compatible with a horizon is reduced
still further. In this sense the rotating extension does not weaken the
super-extremality problem of the large-charge branch, but sharpens it.

A second qualitative effect of rotation concerns the electromagnetic sector itself. A rotating compact object immersed in an external magnetic field develops an induced charge through the Wald mechanism. Parametrically,
\begin{equation}
Q_{\text{ind}} \sim 2 B J,
\label{eq:Wald_charge}
\end{equation}
where $B$ is the characteristic magnetic field strength. This provides a dynamical mechanism for generating a nonzero charge even if the initial configuration is neutral. More importantly, it introduces a direct coupling between the evolution of the charge and that of the angular momentum. In the minimal capture model, the evolution equation for the charge is therefore modified to
\begin{equation}
\frac{dQ}{dt} = C_0 + C_1 Q + C_J J,
\label{eq:Q_evolution_rot}
\end{equation}
where the additional term encodes the rotationally induced contribution. This term can bias the sign of the charge and may therefore provide a microscopic mechanism for selecting a preferred branch of the evolution.

The capture process itself is also modified. In the static case, particle
motion is governed by a central effective potential depending only on \(r\).
In the Kerr--Newman geometry the motion is axisymmetric rather than central:
the energy and the axial angular momentum are conserved, but the total angular
momentum is not. The relevant equations follow from the Hamilton--Jacobi
separation found by Carter \cite{Carter1968KerrFamily}. To make the structure
transparent, we use geometrized units \(G=c=k_e=1\) in this paragraph; the
derivation is given in Appendix~\ref{app:carter_hj_derivation}. For a
massive species \(i\), define the specific conserved quantities
\begin{equation}
\varepsilon_i\equiv \frac{E_i}{m_i},
\qquad
\ell_i\equiv \frac{L_{z,i}}{m_i},
\qquad
\alpha_i\equiv \frac{q_i}{m_i},
\end{equation}
where \(q_i\) is the particle charge. The Hamilton--Jacobi action may be
written in separated form as
\begin{equation}
S_i
=
-\varepsilon_i t+\ell_i\phi+S_{r,i}(r)+S_{\theta,i}(\theta),
\end{equation}
with the electromagnetic coupling included through
\(\partial_\mu S_i-\alpha_i A_\mu\). The separated equations are
\begin{equation}
\Sigma^2 \dot r^2 = {\cal R}_i(r),
\qquad
\Sigma^2 \dot\theta^2 = \Theta_i(\theta),
\label{eq:KN_separated_motion}
\end{equation}
where the overdot denotes differentiation with respect to proper time for a
massive particle. The radial potential is
\begin{equation}
{\cal R}_i(r)
=
\left[
(r^2+a^2)\varepsilon_i
-a\ell_i
-\alpha_i Qr
\right]^2
-\Delta
\left[
r^2+(\ell_i-a\varepsilon_i)^2+{\cal K}_i
\right],
\label{eq:KN_radial_potential}
\end{equation}
and the polar potential is
\begin{equation}
\Theta_i(\theta)
=
{\cal K}_i
-\cos^2\theta
\left[
a^2(1-\varepsilon_i^2)
+\frac{\ell_i^2}{\sin^2\theta}
\right].
\label{eq:KN_polar_potential}
\end{equation}
Here \({\cal K}_i\) is the Carter constant. In the equatorial plane
\(\theta=\pi/2\), one has \({\cal K}_i=0\). For neutral particles
\(\alpha_i=0\); for a nonrotating source \(a=0\), the dependence on the
orientation of \(L_z\) disappears and one recovers the central RN problem.

Equation~\eqref{eq:KN_radial_potential} displays the three effects relevant for
capture. The combination \((r^2+a^2)\varepsilon_i-a\ell_i\) is the usual
Kerr frame-dragging combination. It distinguishes co-rotating orbits
\((a\ell_i>0)\) from counter-rotating orbits \((a\ell_i<0)\). The term
\(-\alpha_i Qr\) is the electromagnetic work term and has opposite sign for
oppositely charged species. Finally, the \(\Delta\)-term contains the rest-mass
and angular barriers.

The turning points of the radial motion are the zeros of \({\cal R}_i(r)\). A
particle arriving from large radius is captured if there is no outer turning
point outside the surface of the object, or outside the outer horizon when a
horizon exists. The boundary between scattering and capture is obtained from an
unstable spherical orbit,
\begin{equation}
{\cal R}_i(r_c)=0,
\qquad
\frac{d{\cal R}_i}{dr}(r_c)=0,
\qquad
{\color{black}\frac{d^2{\cal R}_i}{dr^2}(r_c)>0}.
\label{eq:KN_capture_boundary}
\end{equation}
Solving these equations gives the critical values of \(\ell_i\) and
\({\cal K}_i\), hence the critical impact parameters of the incoming particle.
Because the result depends separately on \(\ell_i\), \({\cal K}_i\), \(a\), and
the sign of \(\alpha_i Q\), the capture cross section is no longer a single
central quantity. Schematically,
\begin{equation}
\sigma_i
=
\sigma_i(\varepsilon_i,\ell_i,{\cal K}_i;a,Q),
\label{eq:sigma_anisotropic}
\end{equation}
and it is generally enhanced for co-rotating trajectories while being suppressed
for counter-rotating trajectories. Charge further splits the electron and proton
capture domains through the sign of \(\alpha_i Q\).

This modification propagates into the saturation mechanism. In the static analysis,
saturation occurs when Coulomb repulsion balances gravitational attraction for the
species with the same sign as the object's charge, namely the species whose infall
{\begin{reftwo}has lost its gravitational focusing contribution.\end{reftwo}} For the natural positive branch this species is the proton; for
an imposed negative branch it is the electron. The corresponding force-balance
condition may therefore be written as
\begin{equation}
GM m_* \sim k_e |Q| e,
\label{eq:static_saturation}
\end{equation}
with \(m_*=m_p\) on the positive proton-limited branch and \(m_*=m_e\) on the
negative electron-limited branch.

{\begin{reftwo}
In the rotating case there is no single central force-balance condition of this
form.  The source angular momentum \(J\) fixes the Kerr--Newman geometry, but
it does not by itself determine the angular momentum distribution of incident
particles.  The saturation problem must therefore be recomputed from the
capture domains defined by
Eq.~\eqref{eq:KN_capture_boundary}, including the distribution of
\(\ell_i\), \({\cal K}_i\), and the sign of \(\alpha_iQ\).  Rotation may bias
capture through frame dragging and through the Wald-induced charge in
Eq.~\eqref{eq:Wald_charge}, but no universal suppression factor for
\(Q_{\max}\) follows from the equations written here.
\end{reftwo}}

{\begin{reftwo}
Rotation may also introduce an ergoregion, but this is a condition on the
actual metric rather than a consequence of rotation alone.  With the
\((+,-,-,-)\) signature used here, the stationary Killing vector becomes
spacelike when \(g_{tt}<0\).  Possible ergoregion instabilities of a
horizonless rotating completion should therefore be discussed only for
geometries that actually contain such a region.
\end{reftwo}}

Finally, the optical properties are qualitatively modified. The notion of a photon sphere generalizes to a photon region, and the resulting lensing is no longer spherically symmetric. Frame dragging introduces asymmetries in deflection angles and can distort the apparent shadow of the object. In strongly super-extremal configurations, the absence of a photon region leads to a suppression of strong-field lensing, reinforcing the conclusions obtained in the static analysis.

Taken together, these results show that the rotating extension of the astron scenario is governed by a two-parameter family $(\Xi_Q, \Xi_J)$ rather than a single charge parameter. Rotation provides both a mechanism for generating charge and an additional constraint on its maximal value. It also strengthens the conclusion that the large-charge branch corresponds to a deeply super-extremal regime, whose physical realization depends sensitively on the interior completion and on the stability of horizonless configurations. In this sense, the static Reissner--Nordstr\"om analysis should be regarded as a limiting case of a more general Kerr--Newman framework in which charge and rotation are intrinsically coupled.

\section{Lensing Around an Astron Singularity}

If astrons are described by highly charged compact objects whose exterior
geometry is of Reissner--Nordstr\"om type, then their lensing properties are
controlled by null geodesics in
\begin{pubrev}Eq.~\eqref{eq:rn_metric_astron}, evaluated at
\(M=M_A\) and \(Q=Q_A\).\end{pubrev}
This discussion is purely geometrical, but it is closely related to
the radiative lens-equation and post-Newtonian analyses of photon and neutrino
lensing in black-hole backgrounds, as well as to electroweak corrections to
photon scattering and polarization in a gravitational field
\cite{CorianoEtAl2015NeutrinoPhotonLensing,CorianoEtAl2015ElectroweakLensing}.
In the equatorial plane, \(\theta=\pi/2\), the null geodesics satisfy
\begin{equation}
\dot r^2 + V_{\rm eff}(r)=E^2,
\qquad
V_{\rm eff}(r)=\frac{L^2}{r^2}f(r),
\end{equation}
where \(E\) and \(L\) are the conserved photon energy and angular momentum.
The bending angle is therefore determined by the effective potential
\(V_{\rm eff}\), exactly as in the Schwarzschild case, but now modified by the
charge term \(Q_A^2/r^2\).

\begin{figure}[t]
\centering
\includegraphics[width=0.64\textwidth]{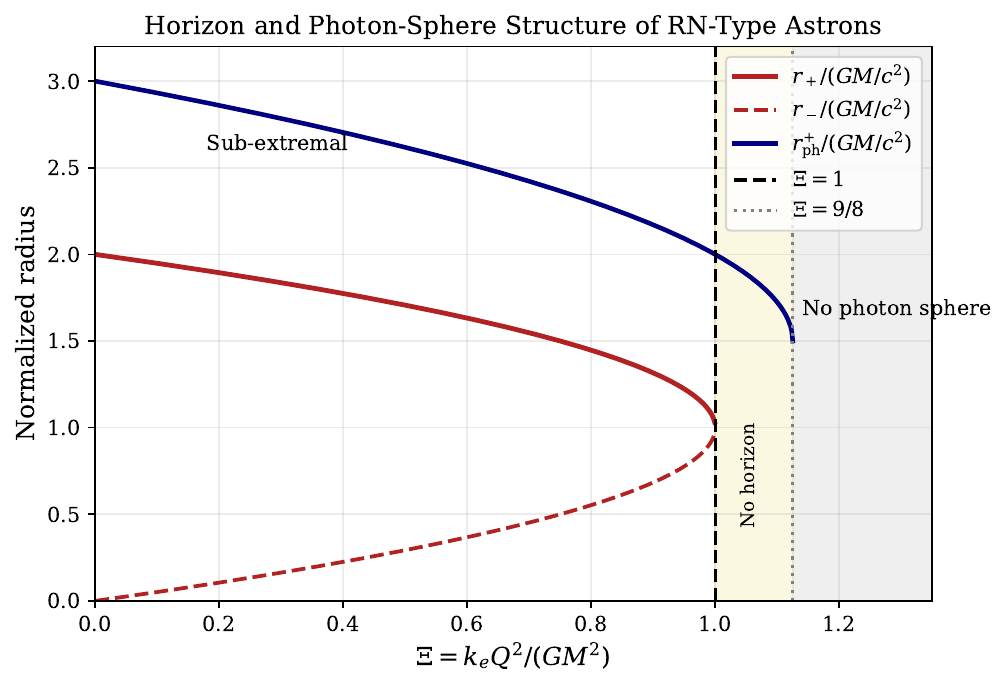}
\caption{Normalized Reissner--Nordstr\"om characteristic radii as functions of
\(\Xi\), defined in Eq.~\eqref{eq:xi_eta_def}. The outer and inner horizons
\(r_\pm\) exist only for
\(\Xi\le1\), where they merge at extremality. The outer photon-sphere radius
survives into the mildly super-extremal regime and disappears only at
\(\Xi=9/8\). {\color{black}The solid blue curve labelled
\(r_{\rm ph}^{+}\) is the outer photon-sphere radius, or equivalently the
radius of the outer unstable circular photon orbit; it is not a horizon radius
or the material radius of the object. The yellow band marks the horizonless interval in
which a photon sphere remains, while the gray band marks the regime without a
photon sphere.} This figure makes transparent why the interval
\(1<\Xi\le9/8\) is lensing-relevant: the object is already horizonless, but it
still possesses a photon sphere.}
\label{fig:rn_radii_vs_xi}
\end{figure}

A first important difference from Schwarzschild appears in the existence of a
photon sphere. Circular null orbits satisfy
\begin{equation}
\frac{d}{dr}\left(\frac{f(r)}{r^2}\right)=0,
\label{eq:photon_sphere_condition}
\end{equation}
which gives
\begin{equation}
r^2 - 3\frac{GM_A}{c^2}r + 2\frac{Gk_eQ_A^2}{c^4}=0.
\end{equation}
Hence the photon-sphere radii are
\begin{equation}
r_{\rm ph}^{\pm}
=
\frac{GM_A}{2c^2}
\left(
3 \pm \sqrt{9-8\Xi}
\right).
\end{equation}
\begin{pubrev}
In the following discussion and in Fig.~\ref{fig:rn_radii_vs_xi},
\(r_{\rm ph}\equiv r_{\rm ph}^{+}\) denotes the outer photon-sphere radius,
namely the radius of the outer unstable circular photon orbit. It is distinct
from the horizon radii \(r_\pm\) and from the material radius \(R\) of the
compact object.
\end{pubrev}
Here \(\Xi\) is the parameter defined in Eq.~\eqref{eq:xi_eta_def}, evaluated
for \(M_A\) and \(Q_A\).
Real photon spheres therefore exist only if
\begin{equation}
\Xi \le \frac{9}{8}.
\end{equation}
By contrast, horizons exist only for
\begin{equation}
\Xi \le 1.
\end{equation}
This leads to three qualitatively different regimes:
\begin{equation}
\Xi<1:
\ \text{charged black hole with horizon and photon sphere},
\end{equation}
\begin{equation}
1<\Xi\le \frac98:
\ \text{naked singularity, but still with a photon sphere},
\end{equation}
\begin{equation}
\Xi>\frac98:
\ \text{naked singularity without a photon sphere}.
\end{equation}

The geometry behind these three regimes is shown explicitly in
Fig.~\ref{fig:rn_radii_vs_xi}. As \(\Xi\) increases from the Schwarzschild
limit, the outer and inner RN horizons move toward each other and merge at
\(\Xi=1\). The outer photon-sphere radius persists slightly beyond extremality
and disappears only at \(\Xi=9/8\). The interval \(1<\Xi\le 9/8\) is therefore
special: it corresponds to a horizonless configuration that still retains a
photon sphere and can consequently mimic part of the strong-lensing structure
of a black hole. Once \(\Xi>9/8\), even this last remnant of black-hole-like
optics is lost.

Here the phrase photon sphere has a precise meaning: it denotes the unstable
circular null orbit, or in a rotating geometry the corresponding photon region,
that underlies the usual black-hole shadow, photon ring and logarithmic
strong-deflection behaviour. Its absence should not be confused with the absence
of gravitational lensing. A massive astron still deflects light through the
ordinary gravitational field, and in the weak-field regime the leading
Schwarzschild-like bending term remains present. What is lost when the photon
sphere disappears is the black-hole-like strong-lensing structure associated with
light orbiting the compact object many times before escaping.

\subsection{Schwarzschild versus extremal Reissner--Nordstr\"om lensing}

A useful elementary comparison is obtained by fixing the mass and comparing the
Schwarzschild metric with the extremal Reissner--Nordstr\"om metric. In
geometrized units, \(G=c=k_e=1\), the line element is
\begin{equation}
ds^2
=
f(r)dt^2-f(r)^{-1}dr^2-r^2d\Omega^2,
\qquad
f(r)=1-\frac{2M}{r}+\frac{Q^2}{r^2}.
\label{eq:rn_metric_lensing_compare}
\end{equation}
Schwarzschild corresponds to \(Q=0\), while extremal RN corresponds to
\(Q^2=M^2\), namely
\begin{equation}
f_{\rm Schw}(r)=1-\frac{2M}{r},
\qquad
f_{\rm ext}(r)=1-\frac{2M}{r}+\frac{M^2}{r^2}
=\left(1-\frac{M}{r}\right)^2 .
\label{eq:schw_ext_f_compare}
\end{equation}
The two metrics have the same \(1/r\) term, but differ at order \(1/r^2\). This
already shows why their lensing can agree at leading weak-field order while
being different beyond leading order.

For equatorial null geodesics, the conserved energy and angular momentum are
\begin{equation}
E=f(r)\dot t,
\qquad
L=r^2\dot\phi,
\label{eq:null_constants_compare}
\end{equation}
and the null condition gives
\begin{equation}
\dot r^2
=
E^2-\frac{L^2}{r^2}f(r).
\label{eq:null_radial_compare}
\end{equation}
Equivalently, with impact parameter \(b=L/E\), the distance of closest approach
\(r_0\) satisfies
\begin{equation}
b^2=\frac{r_0^2}{f(r_0)}.
\label{eq:b_r0_compare}
\end{equation}
The bending angle is therefore
\begin{equation}
\hat\alpha(b)
=
2\int_{r_0}^{\infty}
\frac{b\,dr}{r^2
\sqrt{1-b^2 f(r)/r^2}}
-\pi .
\label{eq:deflection_integral_compare}
\end{equation}
Expanding this integral for \(M/b\ll1\) and \(Q^2/b^2\ll1\) gives
\begin{equation}
\hat\alpha_{\rm RN}(b)
=
\frac{4M}{b}
+\frac{\pi}{4b^2}\left(15M^2-3Q^2\right)
+O(b^{-3}).
\label{eq:weak_deflection_compare}
\end{equation}
Thus the leading term is identical for Schwarzschild and extremal RN,
\begin{equation}
\hat\alpha_{\rm Schw}(b)
=
\frac{4M}{b}
+\frac{15\pi M^2}{4b^2}
+O(b^{-3}),
\label{eq:weak_schw_compare}
\end{equation}
whereas
\begin{equation}
\hat\alpha_{\rm ext}(b)
=
\frac{4M}{b}
+\frac{3\pi M^2}{b^2}
+O(b^{-3}).
\label{eq:weak_ext_compare}
\end{equation}
The equality is therefore only a leading weak-field equality. The charge term
reduces the next-to-leading focusing correction.

The strong-lensing difference is even clearer from the photon sphere. Circular
null orbits are extrema of \(f(r)/r^2\), as stated in
\begin{pubrev}Eq.~\eqref{eq:photon_sphere_condition}.\end{pubrev}
This condition gives
\begin{equation}
r^2-3Mr+2Q^2=0 .
\label{eq:photon_compare_quadratic}
\end{equation}
For Schwarzschild this gives the usual photon sphere
The critical impact parameter follows from the same radial equation. At a
circular null orbit one has \(\dot r=0\), and therefore
\begin{equation}
E^2=\frac{L^2}{r_{\rm ph}^2}f(r_{\rm ph}).
\label{eq:bc_derivation_step}
\end{equation}
Since the impact parameter is \(b=L/E\), Eq.~\eqref{eq:bc_derivation_step}
immediately gives
\begin{equation}
b_c^2=\frac{r_{\rm ph}^2}{f(r_{\rm ph})},
\qquad
b_c=\frac{r_{\rm ph}}{\sqrt{f(r_{\rm ph})}}.
\label{eq:bc_general_static}
\end{equation}
For Schwarzschild this gives
\begin{equation}
r_{\rm ph}^{\rm Schw}=3M,
\qquad
b_c^{\rm Schw}
=
\frac{r_{\rm ph}}{\sqrt{f(r_{\rm ph})}}
=3\sqrt{3}\,M.
\label{eq:schw_photon_bc_compare}
\end{equation}
For extremal RN, \(Q^2=M^2\), the roots are \(r=M\) and \(r=2M\). The external
unstable photon sphere is
\begin{equation}
r_{\rm ph}^{\rm ext}=2M,
\qquad
b_c^{\rm ext}
=
\frac{2M}{\sqrt{f_{\rm ext}(2M)}}
=4M.
\label{eq:ext_photon_bc_compare}
\end{equation}
Thus the critical impact parameter, and hence the shadow scale in this
spherically symmetric comparison, is smaller for extremal RN than for
Schwarzschild at the same mass. The two spacetimes therefore do not have
identical lensing: they agree only in the leading asymptotic weak-field term.

This classification is crucial for lensing. If a photon sphere exists, then the
spacetime can support strong-field lensing with large deflection angles and the
usual sequence of relativistic images associated with photons that loop around
the compact object many times. In that case, a naked singularity can mimic
several of the optical features often associated with black holes. If no photon
sphere exists, however, the lensing structure changes qualitatively: the
logarithmic strong-deflection regime disappears, there is no infinite sequence
of relativistic images, and the object does not possess the standard
photon-sphere shadow characteristic of Schwarzschild or Kerr black holes.
For the benchmark astron parameters one finds
the value in Eq.~\eqref{xi}, namely \(\Xi_A\simeq 5.4\).
This is far above both the horizon threshold \(\Xi=1\) and the photon-sphere
threshold \(\Xi=9/8\). Thus, if one adopts this large-charge benchmark
literally, the corresponding astron would be a strongly super-extremal naked
singularity with no photon sphere. In that regime one should not expect the
standard strong-lensing phenomenology of a black hole. This statement is not a
claim of optical invisibility: the astron mass still produces gravitational
deflection. The lensing would instead be dominated by ordinary weak-field
deflection, with charge corrections that tend to reduce the total bending angle.

Indeed, in the weak-field regime the deflection angle for a Reissner--Nordstr\"om
lens can be expanded as
\begin{equation}
\hat\alpha(b)
\simeq
\frac{4GM_A}{c^2 b}
+
\frac{3\pi}{4b^2}
\left(
5\frac{G^2M_A^2}{c^4}
-
\frac{Gk_eQ_A^2}{c^4}
\right),
\end{equation}
where \(b\) is the impact parameter. The first term is the usual Schwarzschild
bending, while the charge contribution enters with the opposite sign at the next
order and therefore weakens the focusing effect. In this sense, electric charge
acts against gravitational lensing rather than enhancing it.

\begin{figure}[t]
\centering
\includegraphics[width=0.64\textwidth]{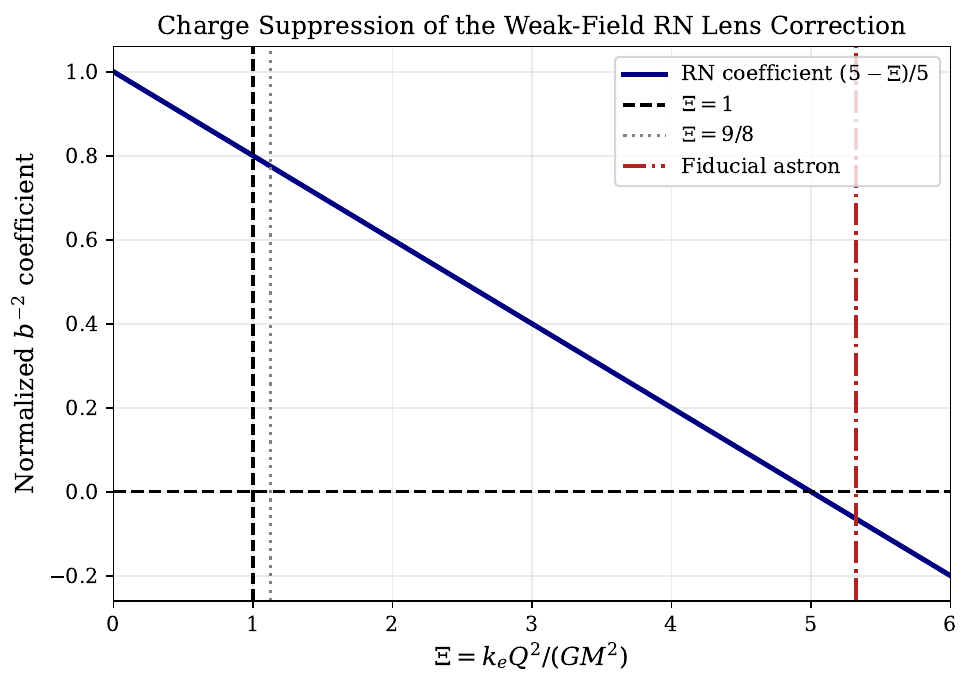}
\caption{Normalized coefficient of the \(b^{-2}\) correction in the weak-field
Reissner--Nordstr\"om bending angle, relative to its Schwarzschild value. The
{\color{black}solid blue curve is the ratio \((5-\Xi)/5\).} The coefficient
decreases linearly with \(\Xi\), showing that electric charge weakens
the next-to-leading focusing term. The vertical markers indicate the extremality
threshold, the photon-sphere threshold, and the fiducial astron value.}
\label{fig:lensing_deflection}
\end{figure}

This suppression can be displayed directly by isolating the coefficient of the
\(b^{-2}\) term. Relative to Schwarzschild, the corresponding RN correction is
reduced by the factor \((5-\Xi)/5\), which decreases linearly with \(\Xi\) and
changes sign for sufficiently large charge. Figure~\ref{fig:lensing_deflection}
shows this normalized coefficient together with the extremality threshold, the
photon-sphere threshold, and the fiducial astron value. The main message is not
that the weak-field expansion remains quantitatively trustworthy arbitrarily close
to the source, but that already at the level of the analytic correction term the
trend is monotonic: increasing charge weakens the second-order focusing effect.

It is useful to compare this behaviour with other naked-singularity families.
The Janis--Newman--Winicour spacetime provides a scalar-field example in which,
depending on the scalar charge, a photon sphere may or may not exist. In this
respect the astron singularity resembles JNW geometries at the level of
lensing phenomenology: both may fall either into a black-hole-like strong-lensing
class or into a weakly focusing class without a photon sphere. The physical
origin, however, is different. In the astron case the suppression of lensing is
driven by electric charge in the Einstein--Maxwell sector, whereas in JNW it is
driven by a scalar field.

The main conclusion is therefore simple. If the astron charge is large enough to
make the object only mildly super-extremal, then strong lensing similar to that
of a black hole may still survive. If, however, one adopts the much larger
charge benchmark in Eq.~\eqref{eq:fiducial_large_charge}, then the object lies
deep in the regime \(\Xi>9/8\), where the photon sphere disappears and the standard
black-hole-like lensing structure is lost. Thus lensing directly measures how far into the super-extremal regime the astron hypothesis is
being pushed.

\section{Lyman-\(\alpha\) Absorption as a Probe of Astron Electric Fields}
\label{sec:lya_field_probe}

The lensing discussion probes the metric produced by an astron.  A different
kind of observable would probe the electric field itself.  The natural
astrophysical system to consider is neutral hydrogen, since the spectra of
high-redshift quasars contain many intervening Lyman-\(\alpha\) absorption
features.  The Lyman-\(\alpha\) forest is produced when quasar light crosses
neutral hydrogen clouds at different redshifts.  The rest-frame transition is
the \(1s\rightarrow 2p\) absorption line of hydrogen,
\begin{equation}
\lambda_\alpha = 1215.67\,\text{\AA},
\end{equation}
so an absorber at redshift \(z_{\rm abs}\) produces a line at
\begin{equation}
\lambda_{\rm obs}
=
\lambda_\alpha(1+z_{\rm abs}).
\end{equation}
Many absorbers along the same line of sight generate the forest of absorption
features observed blueward of the quasar Lyman-\(\alpha\) emission line
\cite{Rauch1998,Meiksin2009}.  Denser neutral systems, such as Lyman-limit
systems and damped Lyman-\(\alpha\) absorbers, provide related probes of neutral
hydrogen with larger column densities and more complicated astrophysical
environments.

The possible connection with astrons is conceptually simple.  Neutral hydrogen
does not experience a net Coulomb acceleration as a charged particle would, but
it is polarizable.  For an absorber \(i\), the observed line would be
schematically
\begin{equation}
\lambda_{{\rm obs},i,\alpha}
=
(1+z_i)
\left[
\lambda_\alpha^{(0)}
+\Delta\lambda_\alpha(E_i)
\right],
\qquad
E_i=E(r_i;Q_A,\lambda_D,\hbox{environment}),
\end{equation}
where \(\alpha\) labels a possible Stark component.  The field-dependent part is
therefore absorber-dependent rather than a universal shift of the whole forest.

For hydrogen, an external electric field perturbs the atomic Hamiltonian by
\begin{equation}
H_{\rm Stark}=e\,{\bf E}\cdot{\bf r},
\end{equation}
or \(H_{\rm Stark}=eEz\) if the field is chosen along the \(z\)-axis.  In the
ideal Coulomb problem, the \(n=2\) level contains \(2n^2=8\) spin states, and
the \(2s\) state mixes linearly with the \(2p,m=0\) state.  The two shifted
linear combinations have
\begin{equation}
\Delta E_{2,\pm}=\pm 3ea_0E,
\end{equation}
while the ground state has no linear Stark shift in the same approximation.
More generally, in parabolic quantum numbers,
\begin{equation}
\Delta E_{n,k}
=
\frac{3}{2}\,n\,k\,e a_0 E,
\label{eq:linear_stark_hydrogen}
\end{equation}
where \(a_0\) is the Bohr radius, \(E=|{\bf E}|\), and \(k\) is the parabolic
quantum-number difference.  Since Lyman-\(\alpha\) absorption involves the
\(n=2\) level, the characteristic frequency scale is
therefore
\begin{equation}
\Delta \nu_2\sim \frac{3e a_0E}{h}.
\label{eq:lya_stark_frequency}
\end{equation}
Equivalently, using \(E_{21}\simeq 10.2\,{\rm eV}\),
\begin{equation}
\left|\frac{\Delta\lambda}{\lambda}\right|
\simeq
\frac{3ea_0E}{E_{21}}
\simeq
1.6\times10^{-11}
\left(\frac{E}{1\,{\rm V\,m^{-1}}}\right),
\end{equation}
or
\begin{equation}
\Delta v
\simeq
4.7\times10^{-3}
\left(\frac{E}{1\,{\rm V\,m^{-1}}}\right)
{\rm m\,s^{-1}}.
\label{eq:lya_velocity_scale_per_field}
\end{equation}
An unscreened astron electric field could therefore in principle induce Stark
shifts, line splittings or excess broadening of neutral-hydrogen absorption
features, but the required field scale is large.

For an isolated astron with charge \(Q_A\), the unscreened field is
\begin{equation}
E_A(r)=\frac{k_eQ_A}{r^2}.
\label{eq:astron_electric_field_unscreened}
\end{equation}
At a megaparsec this gives
\begin{equation}
E_A(1\,{\rm Mpc})
\simeq
3.8\times10^{-3}\,{\rm V\,m^{-1}}
\left(\frac{Q_A}{4\times10^{32}\,{\rm C}}\right).
\label{eq:astron_field_mpc}
\end{equation}
Substituting this value into Eq.~\eqref{eq:lya_stark_frequency} gives only
\begin{equation}
\Delta \nu_2
\sim
1.5\times10^2\,{\rm Hz}
\left(\frac{E_A}{4\times10^{-3}\,{\rm V\,m^{-1}}}\right),
\end{equation}
or a velocity-equivalent shift
\begin{equation}
\Delta v
\sim
c\,\frac{\Delta\nu_2}{\nu_\alpha}
\sim
2\times10^{-5}\,{\rm m\,s^{-1}}.
\end{equation}
This is completely negligible compared with ordinary thermal and turbulent
line widths in the intergalactic medium.  For example, a hydrogen cloud at
\(T\sim10^4\,{\rm K}\) has a thermal Doppler parameter
\begin{equation}
b_{\rm th}
=
\left(\frac{2k_BT}{m_H}\right)^{1/2}
\simeq
13\,{\rm km\,s^{-1}}.
\end{equation}
Thus a megaparsec-scale unscreened field would not produce a directly resolvable
Stark splitting in an ordinary Lyman-\(\alpha\) forest line.

One can invert the estimate to see what field would be needed.  Requiring a
frequency shift comparable to a Doppler width \(b\) gives
\begin{equation}
E_{\rm req}
\sim
\frac{h\nu_\alpha}{3ea_0}\frac{b}{c}
\simeq
2\times10^{6}\,{\rm V\,m^{-1}}
\left(\frac{b}{10\,{\rm km\,s^{-1}}}\right).
\label{eq:lya_required_field}
\end{equation}
For the fiducial charge this unscreened field occurs only within
\begin{equation}
r_{\rm req}
\sim
\left(\frac{k_eQ_A}{E_{\rm req}}\right)^{1/2}
\simeq
40\,{\rm pc}
\left(\frac{Q_A}{4\times10^{32}\,{\rm C}}\right)^{1/2}
\left(\frac{10\,{\rm km\,s^{-1}}}{b}\right)^{1/2}.
\label{eq:lya_required_radius}
\end{equation}
This is much smaller than the typical inter-astron separation and lies in a
region where the gas may be ionized, dynamically disturbed, or otherwise
different from the diffuse neutral absorbers responsible for the usual forest.
The more promising targets would therefore not be generic forest lines far from
the source, but unusually neutral gas associated with the astron environment,
such as high-column-density absorbers or extended gas clouds in which some
neutral fraction survives.

The calculation above gives the natural scale of the effect in the ideal hydrogen
problem. In realistic absorbers the \(n=2\) degeneracy is already split by fine
structure, hyperfine structure and the Lamb shift, while the observed
Lyman-\(\alpha\) profile is shaped by thermal broadening, peculiar velocities,
Hubble flow across the absorber, saturation, radiative transfer and instrumental
resolution \cite{Rauch1998,Draine2011}. For a megaparsec-distance fiducial
astron field, the Stark scale is far below these astrophysical widths. The useful
conclusion is therefore qualitative: generic forest lines are not expected to
resolve a far-field splitting, whereas neutral gas much closer to an unscreened
source could in principle probe the survival of the electric field.

Screening introduces a further suppression.  If the field is Yukawa screened on
a scale \(\lambda_D\), the electrostatic potential is approximately
\begin{equation}
\Phi(r)\simeq \frac{k_eQ_A}{r}e^{-r/\lambda_D},
\end{equation}
and the radial field becomes
\begin{equation}
E(r)
\simeq
\frac{k_eQ_A}{r^2}
\left(1+\frac{r}{\lambda_D}\right)e^{-r/\lambda_D}.
\label{eq:screened_astron_field_lya}
\end{equation}
Thus if \(\lambda_D\ll r\), the neutral hydrogen sees essentially no long-range
electric field.  A Lyman-\(\alpha\) Stark signal would therefore be a direct
probe not only of the charge but also of the survival of the electric field
through the surrounding plasma.  Conversely, the absence of such a signal in
systems expected to pass close to an astron environment would bound the
combination of \(Q_A\), \(\lambda_D\), and the neutral-gas geometry.

The observational role of Lyman-\(\alpha\) absorption is consequently best
understood as a way to constrain the model rather than as an immediate discovery
channel.  A positive signal would require a rare alignment in which neutral
hydrogen lies sufficiently close to an unscreened astron field while remaining
cool and neutral enough to produce identifiable absorption.  A null result,
especially in high-column-density neutral systems near candidate massive dark
seeds, could instead bound the combination of charge, screening length and
neutral gas geometry.  In this sense, Lyman-\(\alpha\) absorption complements
lensing: lensing tests the metric and photon-sphere structure, while
Lyman-\(\alpha\) spectroscopy tests whether the electric field itself survives
into regions containing neutral hydrogen.

\section{Cosmological Implications of the Homogeneous Approximation}

The large-scale dynamics of the Universe in the presence of an astron population
must first be compared with the standard homogeneous FLRW framework.  In units
\(c=1\), the metric is
\begin{equation}
ds^2 = dt^2 - a(t)^2 \left[\frac{dr^2}{1-kr^2} + r^2 d\Omega^2 \right].
\end{equation}
For this metric the Einstein tensor has the components
\begin{equation}
G_{00}=3\left(H^2+\frac{k}{a^2}\right),
\qquad
G^i{}_{j}
=
\left(
2\frac{\ddot a}{a}
 + H^2+\frac{k}{a^2}
\right)\delta^i{}_{j},
\end{equation}
where \(H=\dot a/a\).  A homogeneous and isotropic source is a perfect fluid,
\begin{equation}
T_{\mu\nu}=(\rho+p)u_\mu u_\nu-pg_{\mu\nu},
\qquad
u^\mu=(1,0,0,0).
\end{equation}
Substituting these expressions into Einstein's equations gives the Friedmann
equations
\begin{equation}
H^2 \equiv \left(\frac{\dot a}{a}\right)^2 = \frac{8\pi G}{3}\rho_{\text{tot}} - \frac{k}{a^2},
\end{equation}
\begin{equation}
\frac{\ddot a}{a} = -\frac{4\pi G}{3}(\rho_{\text{tot}} + 3p_{\text{tot}}).
\end{equation}
The Bianchi identity gives the continuity equation
\begin{equation}
\dot\rho+3H(\rho+p)=0.
\end{equation}
Thus a component with \(p=w\rho\) scales as
\begin{equation}
\rho(a)=\rho_0a^{-3(1+w)}.
\end{equation}
Matter, radiation, and a cosmological constant therefore scale as
\begin{equation}
\rho_m\propto a^{-3},
\qquad
\rho_r\propto a^{-4},
\qquad
\rho_\Lambda=\mathrm{const.}
\end{equation}
In the traditional \(\Lambda\)CDM analysis, the observed acceleration is produced
by the last component: \(\rho_\Lambda=\mathrm{const.}\), \(p_\Lambda=-\rho_\Lambda\),
so that \(\rho_\Lambda+3p_\Lambda=-2\rho_\Lambda<0\).  The question for the
astron scenario is whether the charged population can generate an analogous
late-time source, or whether it merely adds ordinary matter and radiation-like
terms.

It is useful to compare this homogeneous treatment with exact Einstein--Maxwell
constructions in which multiple charged black holes are embedded in a cosmological
geometry. Such solutions show explicitly that electric charge can modify the size
of the cosmological region and contribute to backreaction effects, while still allowing
well-defined charged black-hole cosmologies \cite{BibiCliftonDurk2017}. They therefore
support the broader idea that charged compact objects can be incorporated consistently
into cosmological settings, even if they do not by themselves provide late-time
acceleration.

In a homogeneous reduction the astron sector has two conceptually different
pieces.  The first is the rest-mass density of the compact objects.  If the
comoving number density is conserved, then
\begin{equation}
n_A(a)=n_{A0}a^{-3},
\qquad
\rho_{M,A}=M_A n_A\propto a^{-3},
\end{equation}
so the astron rest mass behaves like dust.  The second is the electromagnetic
interaction energy of the charged population.  Thus, schematically,
\begin{equation}
\rho_{\text{tot}}
=
\rho_m+\rho_{M,A}+\rho_C,
\end{equation}
where \(\rho_m\) denotes the non-astron matter sector and \(\rho_C\) denotes the
coarse-grained Coulomb interaction energy.

At this stage it is important to state clearly what this homogeneous treatment
can and cannot establish. If one wishes to reproduce a genuine cosmological-constant
sector within FLRW, one needs an effectively constant energy density,
\begin{equation}
\rho_\Lambda = \text{const.},
\qquad
p_\Lambda=-\rho_\Lambda,
\qquad
w_\Lambda=-1.
\end{equation}
In that case the Friedmann equation acquires an additive constant term and the
late-time expansion approaches de Sitter behavior. The question addressed in this
section is therefore whether the simplest homogeneous reduction of the astron
interaction sector produces anything of that type. As we now show, it does not.

The matter component evolves as usual,
\begin{equation}
\rho_m = \rho_{m0} a^{-3}.
\end{equation}

The homogeneous Coulomb contribution follows from a simple scaling argument.  If
the typical physical separation is \(L(a)=aL_c\), then a pairwise Coulomb energy
scales as
\begin{equation}
V_C(L)\sim \frac{k_eQ_A^2}{L}\propto a^{-1}.
\end{equation}
Since \(n_A\propto a^{-3}\), the interaction energy density scales as
\begin{equation}
\rho_C \sim n_A V_C \propto a^{-4}.
\end{equation}
Equivalently, in a comoving volume \(V\propto a^3\), the total Coulomb energy
scales as \(U_C=\rho_CV\propto a^{-1}\).  The work relation
\(dU_C=-p_CdV\) then gives
\begin{equation}
p_C=\frac{1}{3}\rho_C.
\label{eq:coulomb_pressure}
\end{equation}

The modified Friedmann equation therefore takes the form
\begin{equation}
H^2 = \frac{8\pi G}{3}\left(\rho_{m0} a^{-3} + \rho_{M,A0}a^{-3}+\rho_{C0} a^{-4}\right).
\end{equation}

This scaling law implies that the astron contribution behaves formally as a radiation-like
component at the level of the continuity equation. Its microscopic origin is different
from that of ordinary radiation, but its homogeneous redshift behaviour is the same.
In particular, the interaction term is not constant in time. For that reason alone
it cannot be identified with a cosmological constant in the strict FLRW sense.

To characterize the effective equation of state, one considers the conservation equation
\begin{equation}
\dot{\rho}_C + 3H(\rho_C + p_C) = 0.
\end{equation}
\begin{pubrev}
Substituting \(\rho_C \propto a^{-4}\) reproduces
Eq.~\eqref{eq:coulomb_pressure}.
\end{pubrev}

At the level of the homogeneous background, the astron contribution therefore corresponds to an effective equation-of-state parameter
\begin{equation}
w_C = \frac{1}{3}.
\end{equation}

This result is already highly constraining. Although the Coulomb interaction is repulsive,
the associated homogeneous energy density redshifts as $a^{-4}$ and therefore does
not mimic a cosmological constant. In particular, it does not generate a constant
source term in the Friedmann equation and does not provide a de Sitter-like late-time
solution. This is the precise sense in which the homogeneous astron sector does not,
by itself, justify the cosmological constant.

To make contact with observations, it is useful to rewrite the Friedmann equation in terms of redshift $z = a^{-1} - 1$,
\begin{equation}
H^2(z) = H_0^2 \left[\Omega_{m0}^{\rm eff}(1+z)^3 + \Omega_{C0}(1+z)^4 \right],
\end{equation}
where \(\Omega_{m0}^{\rm eff}\) includes ordinary nonrelativistic matter and the
astron rest-mass contribution.

This should be compared with the standard $\Lambda$CDM expression,
\begin{equation}
H^2_{\Lambda\text{CDM}}(z) = H_0^2 \left[\Omega_{m0}(1+z)^3 + \Omega_{\Lambda}\right].
\end{equation}

The two models differ sharply in their late-time behaviour. In $\Lambda$CDM, the
constant term $\Omega_{\Lambda}$ dominates asymptotically, leading to exponential
expansion,
\begin{equation}
a(t) \sim e^{H_\Lambda t}.
\end{equation}

In contrast, in the homogeneous astron approximation the interaction term decays
as $(1+z)^4$, implying that it becomes subdominant at sufficiently late times. The
expansion therefore returns asymptotically to matter domination,
\begin{equation}
a(t) \propto t^{2/3}.
\end{equation}

This distinction is not merely quantitative but qualitative. A radiation-like term
may temporarily contribute to the energy budget, but it cannot act as a constant
vacuum component. Therefore, even if one normalizes \(\rho_A\) so that it is comparable
to the matter density at some epoch, the model still lacks the constant late-time
driver that characterizes \(\Lambda\)CDM.

Equivalently, the homogeneous astron interaction cannot provide an asymptotic
acceleration era. A component with \(\rho_C\propto a^{-4}\) becomes less important
relative to matter as the universe expands and cannot approach a de Sitter fixed
point. Thus, even in a phenomenological interpretation in which the charged sector
affects the expansion over some finite interval, that effect is transitory unless
additional physics beyond the homogeneous perfect-fluid reduction is supplied.

The present epoch may correspond to a regime in which the Coulomb contribution is
comparable to the matter density,
\begin{equation}
\rho_C \sim \rho_m,
\end{equation}
but that fact alone does not imply acceleration.

The deceleration parameter is
\begin{equation}
q = -\frac{\ddot a}{aH^2} = \frac{1}{2}\left(1 + 3\frac{p_{\text{tot}}}{\rho_{\text{tot}}}\right).
\end{equation}

Substituting the components yields
\begin{equation}
q(z) = \frac{1}{2}\frac{\Omega_{m0}^{\rm eff}(1+z)^3 + 2\Omega_{C0}(1+z)^4}{\Omega_{m0}^{\rm eff}(1+z)^3 + \Omega_{C0}(1+z)^4}.
\end{equation}

For positive densities, the expression above is always positive. The homogeneous
perfect-fluid reduction presented here therefore does \emph{not} generate accelerated
expansion, and in particular it cannot generate asymptotically late-time
acceleration. Instead, it shows that the simplest FLRW treatment of the astron
interaction sector behaves like an additional radiation-like component. This is
the most direct way to formulate the limitation emphasized above: within
homogeneous FLRW, astrons do not produce a constant solution and therefore do not
reproduce dark energy in the form of a cosmological constant. {\begin{reftwo}
Within the equations of this homogeneous model, the result is stronger: for
positive matter and Coulomb energy densities one has \(q>0\), so the
interaction sector does not produce acceleration at any epoch.  A transient
acceleration would be a separate possibility only after additional
inhomogeneous or backreaction physics is specified.\end{reftwo}}

{\begin{refone}
There is a second limitation, independent of the \(a^{-4}\) scaling, which
addresses the physical meaning of acceleration in this setting.  The Coulomb
force is a force on charged sources.  It can generate peculiar accelerations of
astrons with respect to one another, but neutral galaxies, neutral dark matter
and photons are not directly accelerated by the electric field.  They can feel
the charged sector only through the metric, or indirectly if they are
gravitationally bound to structures whose motion is controlled by the astrons.
Thus the question is not whether two same-sign charged objects repel.  The
question is whether the Einstein--Maxwell system generated by the charged
population produces an effective expansion of the common cosmic flow, including
the no-net-charge component.  In the homogeneous reduction just derived, the
answer is negative: the metric source associated with the Coulomb energy is
radiation-like and decelerating.
\end{refone}}

This observation should be read as a limitation, not as a failure of the broader
idea. A cosmological implementation based on discrete charged objects may require
physics beyond a homogeneous perfect-fluid description, for example genuinely inhomogeneous
dynamics, non-perturbative backreaction, or a different coarse-graining prescription.
However, none of those effects is derived in the present paper. What can be stated
firmly is that the simplest background treatment is insufficient to replace
\(\Lambda\)CDM, and that any stronger cosmological claim must rely on physics
beyond the homogeneous approximation.

\section{Backreaction as an Open Problem}

The failure of the homogeneous FLRW reduction does not automatically rule out the
astron framework. It does, however, identify the central open problem. The astron
scenario is intrinsically one of a sparse population of discrete, extremely massive,
charged compact objects interacting through long-range gravitational and electromagnetic
fields. Such a system is not obviously equivalent to a homogeneous perfect fluid,
and reducing it to one may erase precisely the effects one would hope to retain.
The natural place to revisit the cosmology is therefore the backreaction problem:
whether averaging the Einstein--Maxwell dynamics of an inhomogeneous astron population
can lead to an effective large-scale evolution that differs significantly from the
naive FLRW result.

In cosmology, backreaction refers to the fact that averaging and time evolution need
not commute in a nonlinear theory such as general relativity. This issue has been
discussed extensively in the literature, both from the point of view of constructive
averaging formalisms and from the point of view of mathematically controlled
limitations
\begin{pubrev}\cite{Buchert2000,ClarksonEtAl2011,GreenWald2011}\end{pubrev}.
The central question is whether
small- or intermediate-scale inhomogeneities can feed back into the effective expansion
of the Universe strongly enough to mimic, enhance, or otherwise modify the role
that would conventionally be assigned to dark energy.

{\begin{refthree}
The geometric congruence normal to a chosen foliation and the material
congruence of the charged sources need not coincide.  In a flow-orthogonal
Gaussian-normal slicing with unit lapse and zero shift, the hypersurface-normal
congruence is geodesic, whereas the charged matter congruence is accelerated by
the Lorentz force.  The latter must therefore be described by a distinct
four-velocity, or by a tilted congruence relative to the hypersurface normal.
For the signature \((+,-,-,-)\) and \(u_\mu u^\mu=1\), the acceleration
divergence of a nongeodesic congruence satisfies
\begin{equation}
\nabla_\mu a^\mu=D_\mu a^\mu-a_\mu a^\mu,
\end{equation}
so its contribution is not represented by a spatial boundary flux alone.

An averaged Einstein--Maxwell description must consequently distinguish the
foliation normal, the charged-matter velocity, and the velocity of any neutral
component.  Its scalar evolution system also depends on the lapse, relative
tilt, acceleration-squared term, electromagnetic stresses, shear, and boundary
data, together with the corresponding averaging commutation relations.  In
the absence of such a closed system, the sign and magnitude of a cosmological
backreaction cannot be inferred from the pairwise force or lattice-energy
estimates.  The backreaction of a discrete charged population therefore
remains an open Einstein--Maxwell problem.
\end{refthree}}

To make the remaining, purely kinematic estimates concrete, imagine placing
astrons on a cubic lattice with comoving spacing \(L_c\), carried by a
prescribed background scale factor \(a(t)\).  The physical separation is
\begin{equation}
L(t)=a(t)L_c.
\end{equation}
If a finite region \({\cal D}\) contains \(N\) lattice cells, then
\begin{equation}
V_{\cal D}(t)\simeq N a^3 L_c^3,
\qquad
n_A(t)=\frac{N}{V_{\cal D}(t)}
=\frac{1}{a^3L_c^3}.
\end{equation}
The rest-mass density of the astron lattice therefore behaves as dust,
\begin{equation}
\rho_M
=
M_A n_A
=
\frac{M_A}{L_c^3}a^{-3}.
\end{equation}

The Coulomb contribution is obtained by summing the pairwise interaction energy,
\begin{equation}
U_C({\cal D})
=
\frac{1}{2}\sum_{i\neq j\in{\cal D}}
\frac{k_e Q_A^2}{r_{ij}},
\qquad
r_{ij}=aL_c|\mathbf n_i-\mathbf n_j|.
\end{equation}
Equivalently,
\begin{equation}
U_C({\cal D})
=
\frac{k_eQ_A^2}{aL_c}\,{\cal S}_{\cal D},
\end{equation}
where \({\cal S}_{\cal D}\) is the dimensionless lattice sum.
{\begin{reftwo}
For a finite same-sign domain, \({\cal S}_{\cal D}\) is not automatically
proportional to the number of cells \(N\).  A roughly
spherical unscreened same-sign population at fixed spacing has
\(R_{\cal D}\propto N^{1/3}\) and
\(U_C\sim k_e(NQ_A)^2/R_{\cal D}\), hence \(U_C\propto N^{5/3}\).  The
extensive scaling \({\cal S}_{\cal D}\propto N\) follows only after a
screening, compensation, periodic-neutralization, or finite-boundary
prescription has been specified.\end{reftwo}} In that conditional case
\begin{equation}
\rho_C({\cal D})
=
\frac{U_C({\cal D})}{V_{\cal D}}
\propto
a^{-4}.
\end{equation}
Thus a perfectly homogeneous or regularly coarse-grained lattice reproduces the
radiation-like scaling found in the previous section. The lattice picture therefore
does not by itself solve the acceleration problem.

\begin{figure}[t]
\centering
\includegraphics[width=0.54\textwidth]{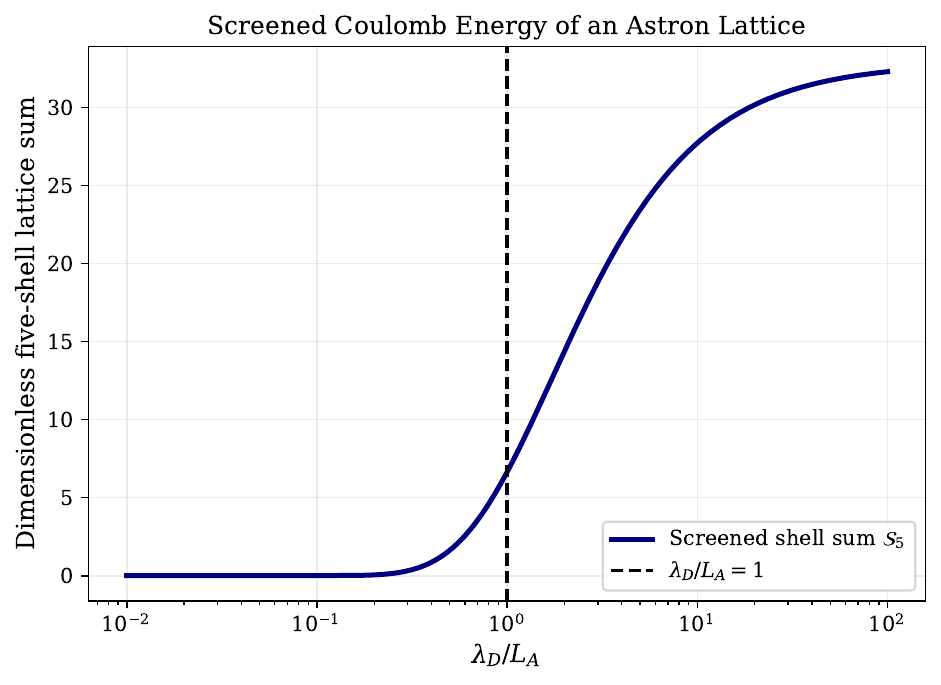}
\caption{Shell-truncated cubic-lattice sum for a Yukawa-screened inter-astron
interaction, shown as a function of \(x=\lambda_D/L_A\). {\color{black}The solid
blue curve is the dimensionless five-shell sum
\({\cal S}_5(x)=\sum_{j=1}^{5}N_j e^{-d_j/x}/d_j\), with shell multiplicities
\(N_j=(6,12,8,6,24)\) and dimensionless radii
\(d_j=(1,\sqrt{2},\sqrt{3},2,\sqrt{5})\).} For \(\lambda_D\ll L_A\)
the interaction is strongly suppressed, while for \(\lambda_D\gtrsim L_A\) an
extended lattice contribution survives. The figure therefore provides a simple
numerical proxy for the interplay between screening and the backreaction problem.}
\label{fig:lattice_screening}
\end{figure}

The role of screening in this lattice picture can be illustrated by replacing the
bare Coulomb interaction by a Yukawa form and evaluating a shell-truncated cubic
lattice sum. Figure~\ref{fig:lattice_screening} shows the resulting normalized
interaction energy as a function of \(\lambda_D/L_A\), where \(L_A\) is the
lattice spacing. For \(\lambda_D\ll L_A\) the interaction is strongly suppressed
and the lattice behaves effectively locally, whereas long-range correlations survive
only once \(\lambda_D\gtrsim L_A\). This is a compact way to visualize why the
screening problem and the backreaction problem are inseparable: if the Debye length
is too short compared with the inter-astron separation, the long-range lattice
interaction needed for any collective cosmological effect is removed from the start.

{\begin{refthree}
A regular lattice reproduces the homogeneous radiation-like scaling and does
not establish acceleration.  Departures from regularity would generate
inhomogeneous expansion, shear, curvature, Maxwell stresses, and relative
motion among charged and neutral components, but the preceding observations do
not determine their averaged sign or magnitude.  The lattice kinematics alone
therefore supply no acceleration condition.
\end{refthree}}

This is also the precise sense in which the astron proposal is different from simply
assigning a scalar pressure to a homogeneous fluid. In a fluid description the
interaction is compressed into \(\rho\) and \(p\). In the discrete Einstein--Maxwell
description one must instead keep the vector Lorentz force, the traceless Maxwell
stress, and the domain dependence of the fields. A further complication is that an
infinite same-sign Coulomb lattice is not globally well defined without a screening
prescription, a finite-domain prescription, or some compensating background. This is
another way in which the screening problem and the cosmological backreaction problem
are intertwined.

Let us now apply these estimates to a fiducial astron lattice using the mass in
Eq.~\eqref{eq:fiducial_astron_mass} and the charge in
Eq.~\eqref{eq:fiducial_large_charge}. Denote the present physical lattice spacing
by \(L_A\). The number density is
\begin{equation}
n_A=L_A^{-3}
\simeq
3.4\times10^{-68}
\left(\frac{L_A}{1\,\mathrm{Mpc}}\right)^{-3}
\mathrm{m}^{-3}.
\end{equation}
The corresponding mass density is
\begin{equation}
\rho_M=M_A n_A
\simeq
6.8\times10^{-26}
\left(\frac{M_A}{10^{12}M_\odot}\right)
\left(\frac{L_A}{1\,\mathrm{Mpc}}\right)^{-3}
\mathrm{kg\,m^{-3}}.
\end{equation}
Using \(H_0\simeq67.4\,\mathrm{km\,s^{-1}\,Mpc^{-1}}\), this corresponds to
\begin{equation}
\Omega_A
=
\frac{\rho_M}{\rho_{\mathrm{crit}}}
\simeq
7.9
\left(\frac{M_A}{10^{12}M_\odot}\right)
\left(\frac{L_A}{1\,\mathrm{Mpc}}\right)^{-3}.
\end{equation}
Thus a literal spacing \(L_A\simeq1\,\mathrm{Mpc}\) would overclose the Universe.
If one wants the astron rest mass to be of order the matter abundance,
\(\Omega_A\sim0.3\), the corresponding spacing is instead
\begin{equation}
L_A\simeq 3\,\mathrm{Mpc}
\left(\frac{M_A}{10^{12}M_\odot}\right)^{1/3}
\left(\frac{0.3}{\Omega_A}\right)^{1/3}.
\end{equation}
For a sparser population, such as \(N_A\sim10^{11}\) objects in a Hubble-scale
volume of radius \(14\,\mathrm{Gpc}\), the mean separation is of order
\begin{equation}
L_A\simeq 5\,\mathrm{Mpc},
\end{equation}
corresponding to \(\Omega_A\sim 0.07\).

The charge benchmark is large enough that the electromagnetic force between two
astrons exceeds their mutual Newtonian attraction, as clear from \eqref{xi}.

This number is independent of the lattice spacing. The acceleration of one astron
due to a single nearest neighbour is instead scale dependent:
\begin{equation}
a_{\mathrm{pair}}
\sim
\frac{k_eQ_A^2-GM_A^2}{M_A L_A^2}
\simeq
6.2\times10^{-13}
\left(\frac{L_A}{1\,\mathrm{Mpc}}\right)^{-2}
\mathrm{m\,s^{-2}}.
\end{equation}
This estimate should not be confused with a cosmological acceleration, since a
symmetric lattice cancels most vector forces by symmetry. Its role is only to show
the local scale of the inter-astron repulsion.

The same point appears at the level of the interaction energy. A nearest-neighbour
estimate gives
\begin{equation}
\rho_C
\sim
\frac{k_eQ_A^2/L_A}{L_A^3}
\simeq
1.6\times10^{-15}
\left(\frac{Q_A}{4\times10^{32}\,\mathrm{C}}\right)^2
\left(\frac{L_A}{1\,\mathrm{Mpc}}\right)^{-4}
\mathrm{J\,m^{-3}}.
\end{equation}
Compared with the rest-mass energy density,
\begin{equation}
\rho_Mc^2
\simeq
6.1\times10^{-9}
\left(\frac{M_A}{10^{12}M_\odot}\right)
\left(\frac{L_A}{1\,\mathrm{Mpc}}\right)^{-3}
\mathrm{J\,m^{-3}},
\end{equation}
one obtains
\begin{equation}
\frac{\rho_C}{\rho_Mc^2}
\sim
2.6\times10^{-7}
\left(\frac{Q_A}{4\times10^{32}\,\mathrm{C}}\right)^2
\left(\frac{M_A}{10^{12}M_\odot}\right)^{-1}
\left(\frac{L_A}{1\,\mathrm{Mpc}}\right)^{-1}.
\end{equation}
Therefore, if the Coulomb sum is effectively local or screened, its energy density
is much too small to act as a dark-energy component. More generally, a screened
lattice would involve the replacement
\begin{equation}
U_C({\cal D};\lambda_D)
=
\frac{1}{2}\sum_{i\neq j\in{\cal D}}
\frac{k_e Q_A^2}{r_{ij}}e^{-r_{ij}/\lambda_D}.
\end{equation}
If \(\lambda_D\ll L_A\), inter-cell correlations are exponentially suppressed and
the backreaction problem becomes essentially local. If instead
\(\lambda_D\gtrsim L_A\), the fields correlate many lattice cells and the averaging
domain \(V_{\cal D}\) must be chosen large enough to contain the relevant nonlocal
electromagnetic stresses. If the same-sign Coulomb field is assumed to remain
unscreened across many lattice cells, the lattice sum is no longer a simple local
quantity and a global prescription is needed. This is precisely why the screening
problem cannot be separated from the cosmological backreaction problem.

{\begin{refthree}
The lattice estimates establish two limited results.  First, a screened or
effectively local Coulomb interaction has an energy density far below the
rest-mass energy density for the benchmark parameters.  Second, extending an
unscreened same-sign field over many cells requires a global prescription for
screening, compensation, or boundary conditions.  Neither result supplies an
averaged acceleration law.  Determining whether the full inhomogeneous
Einstein--Maxwell system changes the expansion of the common charged and
neutral cosmic flow requires a separate, internally consistent averaging or
numerical-relativity calculation.  The magnitude and sign of such a
backreaction remain undetermined.
\end{refthree}}

\section{Conclusions}

The main result of this analysis is not the construction of a viable astron
population, but the identification of the conditions that such a population
would have to satisfy.  These conditions are severe.

First, the minimal capture model does not generate the fiducial charge.  For
\(M_A\sim10^{12}M_\odot\), the linear capture model has a current-balance
fixed point \(Q_{\rm eq}^{(+)}\simeq3.6\times10^{12}\,{\rm C}\), while the
species-dependent focusing-cancellation scales span approximately
\(10^{11}\)--\(10^{14}\,{\rm C}\).  All of these are far below the
phenomenological value \(Q_A\sim4\times10^{32}\,{\rm C}\), leaving a gap of
18--21 orders of magnitude.  Thus the fiducial large-charge branch must be
regarded as a primordial charge-concentration hypothesis, not as the endpoint of
the capture mechanism studied here.

Second, charge persistence has not been established by the failure of linear
Debye screening.  The benchmark charge produces an electrostatic potential for
which \(e\Phi/(k_BT)\gg1\) on the relevant scales, so the linear
Debye--H\"uckel approximation is outside its domain of validity.  This does not
mean that screening or neutralization is ineffective.  It means that the problem
is nonlinear and kinetic: one must show that electron inflow, trapping, drift,
diffusion, and phase-space rearrangement do not neutralize the source on the
timescale of interest.

Third, a cosmologically relevant abundance of \(10^{12}M_\odot\) compact objects
is a large-scale-structure problem.  If astrons carry a fraction \(f_A\) of the
matter density, their mean separation is of order
\[
d_A\simeq
3\,f_A^{-1/3}
\left(\frac{M_A}{10^{12}M_\odot}\right)^{1/3}{\rm Mpc}.
\]
A literal one-megaparsec spacing would overproduce the cosmic matter density,
while a viable sparse population must satisfy constraints from Poisson power,
halo abundances, clustering, lensing and dynamical effects.

The geometric analysis remains useful within this limited interpretation.  The
fiducial large-charge branch lies in the super-extremal
Reissner--Nordstr\"om regime and, for the benchmark parameters, beyond the
photon-sphere threshold.  This affects its lensing phenomenology and clarifies
the distinction between an exact naked singularity and a regular horizonless
charged compact object with the same exterior data.  These statements are
conditional on the existence of the charged object; they do not provide a
formation mechanism.

The cosmological conclusion is similarly restrictive.  In a homogeneous FLRW
reduction, the Coulomb interaction energy of a charged population redshifts as
\(a^{-4}\), so it behaves like radiation rather than a cosmological constant.
{\begin{refone}
Nor does the electric field directly accelerate the no-net-charge component of
the Universe.  Neutral galaxies, photons and neutral dark matter can respond
only through the metric or through gravitational binding to charged structures.
Any late-time accelerating effect would therefore have to come from additional
physics, such as a controlled inhomogeneous Einstein--Maxwell averaging
mechanism shared by the neutral cosmic flow.
\end{refone}}
{\begin{refthree}No definite sign or magnitude for such a backreaction follows
from the kinematic and energy estimates above.\end{refthree}}

Taken together, these results recast the astron proposal as a falsifiable set of
consistency tests.  A successful model must provide a primordial mechanism for
the large charge, demonstrate nonlinear survival against plasma neutralization,
{\begin{refthree}derive, within a consistent dynamical framework, a metric-level
effect on the common cosmic flow,\end{refthree}} and satisfy the
abundance and discreteness constraints associated with ultra-massive compact
objects.  Without these ingredients, the fiducial astron should be treated as a
benchmark for testing, not as an established physical population.

\vspace{0.5cm}
\centerline{\bf Funding}
\begin{pubrev}
This research was funded by the Istituto Nazionale di Fisica Nucleare (INFN)
through the iniziativa specifica \emph{QG-sky}.
\end{pubrev}

\vspace{0.5cm}
\centerline{\bf Acknowledgements} 
C.C. thanks the Yang Institute for Theoretical Physics and the Simons Center at
Stony Brook for hospitality while completing the project.

\appendix

\section{Compact Charging and Free-Fall Benchmarks}
\label{app:tff}

\begin{pubrev}
The exact time dependence of the capture model is given by
Eq.~\eqref{eq:charge_solution_general}, while its response time is given in
Eq.~\eqref{eq:tau_response}. For the normalization used in the numerical
estimates, the latter becomes
\end{pubrev}
\begin{equation}
\tau \simeq
1.6\times10^{1}
\left(\frac{T}{10^6\,{\rm K}}\right)^{1/2}
\left(\frac{R}{1.48\times10^{15}\,{\rm m}}\right)^{-1}
\left(\frac{n}{10^{-13}\,{\rm m}^{-3}}\right)^{-1}
{\rm s}.
\end{equation}
These values should be read only as local response estimates of the linear
capture coefficient.  When the formal value falls below microscopic plasma
times, the macroscopic capture model has left its domain of literal validity.

\begin{pubrev}
For comparison, the free-fall time of a pressureless uniform sphere of density
\(\rho\) is given by Eq.~\eqref{eq:tff_env_comp}.
\end{pubrev}
The comparison between \(\tau\) and \(t_{\rm ff}\) is therefore useful as a
diagnostic of whether local charge rearrangement is faster than gravitational
collapse, but it is not a nonlinear charge-survival calculation.

\section{Origin of the Phenomenological Large-Charge Branch}
\label{app:qm_extrapolation}

The large fiducial value \(Q_A\sim4\times10^{32}\,{\rm C}\) is not obtained from
the minimal capture model.  It follows from the phenomenological extrapolation
used in Ref.~\cite{Frampton2022EAU}.  The assumed input is that the
charge-to-mass ratio grows approximately linearly with mass in logarithmic
variables.  Writing \(M=10^m\,{\rm kg}\), the two benchmark points used in that
argument imply
\begin{equation}
\log_{10}(Q/M)=m-52.
\end{equation}
Therefore
\begin{equation}
Q=M\frac{Q}{M}=10^m10^{m-52}=10^{2m-52}\,{\rm C},
\end{equation}
so
\begin{equation}
Q\propto M^2.
\end{equation}
For \(M_A=10^{12}M_\odot\simeq2\times10^{42}\,{\rm kg}\), this gives
\begin{equation}
Q_A\simeq10^{-52}\left(2\times10^{42}\right)^2\,{\rm C}
\simeq4\times10^{32}\,{\rm C}.
\end{equation}
This appendix is included to make the status of the red curve in
Fig.~\ref{fig:charge_scales} explicit: it is a phenomenological extrapolation,
not the output of the accretion calculation.

\section{Linear Debye Benchmark and Its Domain of Validity}
\label{debye}

For a multicomponent quasi-neutral plasma in weak electrostatic potential
\(\Phi\), the Boltzmann response gives
\begin{equation}
n_s(\Phi)
\simeq
n_{s0}\left(1-\frac{q_s\Phi}{k_BT_s}\right),
\end{equation}
provided \(|q_s\Phi|\ll k_BT_s\).  Substitution into Poisson's equation gives
\begin{equation}
\nabla^2\Phi=\frac{\Phi}{\lambda_D^2}.
\end{equation}
\begin{pubrev}
Here the multicomponent Debye length is defined in
Eq.~\eqref{eq:debye_multicomponent}, and the spherically symmetric point-charge
solution is the Yukawa potential in Eq.~\eqref{eq:debye_yukawa_potential}.
\end{pubrev}
For a symmetric electron--proton plasma with \(n_e=n_p=n_0\) and
\(T_e=T_p=T\),
\begin{pubrev}the result reduces to Eq.~\eqref{eq:debye_two_species}.\end{pubrev}
If the ions are treated as a fixed neutralizing background, one instead obtains
\begin{pubrev}the electron Debye length in
Eq.~\eqref{eq:debye_electron}.\end{pubrev}
Thus the factor of two is a convention about which species are mobile in the
linear response.

For the astron benchmark the expansion parameter
\begin{equation}
\frac{e\Phi}{k_BT}
\end{equation}
is enormous on parsec and megaparsec scales.  Therefore the linear Debye result
is a benchmark for the ambient plasma, not a self-consistent screening solution
for the astron.  Its failure does not imply charge survival; it means that the
problem has become nonlinear.

\section{Minimal Nonlinear Kinetic Screening Problem}
\label{app:kinetic_screening}

A self-consistent neutralization calculation must evolve the charged particle
distribution functions rather than insert the linear Debye length by hand.  In
the simplest electrostatic approximation one may write
\begin{equation}
\frac{\partial f_s}{\partial t}
\;{\color{black}+}\;\mathbf v\cdot\nabla_{\mathbf x}f_s
\;{\color{black}+}\;\frac{q_s}{m_s}
\left(-\nabla\Phi\right)\cdot\nabla_{\mathbf v}f_s
=
C_s[f],
\end{equation}
together with
\begin{equation}
\nabla^2\Phi
=
-\frac{1}{\epsilon_0}
\left[
\rho_{\rm source}
+
\sum_s q_s\int f_s\,d^3v
\right].
\end{equation}
The collision operator \(C_s[f]\) may be set to zero in a collisionless
Vlasov--Poisson approximation or replaced by a Fokker--Planck operator when
collisions are important.  The survival question is then whether the solution
develops a neutralizing electron cloud or current on a timescale shorter than
the cosmological or dynamical time of interest.  Possible bottlenecks include
angular momentum barriers, magnetic fields, trapping, phase-space depletion, and
finite charge reservoirs.  These effects must be demonstrated in a nonlinear
calculation; they do not follow from the failure of linear Debye theory.

\section{Kerr--Newman Separation: Minimal Formulae}
\label{app:carter_hj_derivation}

The rotating discussion uses only the standard separability of charged-particle
motion in Kerr--Newman spacetime.  In geometrized units \(G=c=k_e=1\),
\begin{equation}
\Delta=r^2-2Mr+a^2+Q^2,
\qquad
\Sigma=r^2+a^2\cos^2\theta.
\end{equation}
For a particle of charge \(e\), energy \(E\), azimuthal angular momentum \(L\),
and Carter constant \({\cal K}\), the radial potential can be written as
\begin{equation}
{\cal R}(r)
=
\left[(r^2+a^2)E-aL-eQr\right]^2
-
\Delta\left[\mu^2r^2+(L-aE)^2+{\cal K}\right].
\end{equation}
\begin{pubrev}
The capture boundary is determined by the unstable-orbit conditions in
Eq.~\eqref{eq:KN_capture_boundary}, with the species label suppressed.
\end{pubrev}
The full Hamilton--Jacobi derivation is standard and is not repeated here.

\end{document}